\documentclass{article}

\usepackage{amsmath}

\usepackage{times}
\usepackage{bm}
\usepackage{natbib}
 
 \usepackage{multicol}

\usepackage[plain,noend]{algorithm2e}

\allowdisplaybreaks

\usepackage{latexsym}
\usepackage{multirow}\usepackage{bm}
\usepackage{url}
\usepackage{setspace}
\usepackage{color}
\usepackage{amsmath}
\usepackage{amssymb}
\usepackage{mathtools}
\usepackage{graphicx}
\usepackage{wasysym}
\usepackage{tikz}
\usetikzlibrary{positioning,shapes.geometric,graphs}
\usetikzlibrary[graphs]
 \usepackage{multirow}
\usepackage{multicol}
 \usepackage{hyperref}
\usepackage{float}
\usepackage{comment}
 \usepackage{bbm}
\usepackage{lipsum}
  \usepackage{color}
\usepackage{graphicx,subcaption} 
\usepackage{mathrsfs}
\usepackage{lipsum}
\usepackage{colortbl}
\usepackage{subcaption}
\usepackage{dcolumn}
\usepackage{rotating}

\newcommand{\bone}{\mathbf{1}}

\newcommand{\pr}{pr}

\newcommand{\E}{E}

\usepackage{amssymb}

\usepackage{authblk}

\newcommand{\T}{\mathrm{\scriptscriptstyle T}}

\newcommand{\de}{\rm{d}}

 \newcommand\smallO{
  \mathchoice
    {{\scriptstyle\mathcal{O}}}
    {{\scriptstyle\mathcal{O}}}
    {{\scriptscriptstyle\mathcal{O}}}
    {\scalebox{.7}{$\scriptscriptstyle\mathcal{O}$}}
  }
 
\makeatletter
\renewcommand{\algocf@captiontext}[2]{#1\algocf@typo. \AlCapFnt{}#2} 
\def\@algocf@capt@plain{top}
\renewcommand{\algocf@makecaption}[2]{%
  \addtolength{\hsize}{\algomargin}%
  \sbox\@tempboxa{\algocf@captiontext{#1}{#2}}%
  \ifdim\wd\@tempboxa >\hsize
    \hskip .5\algomargin%
    \parbox[t]{\hsize}{\algocf@captiontext{#1}{#2}}
  \else%
    \global\@minipagefalse%
    \hbox to\hsize{\box\@tempboxa}
  \fi%
  \addtolength{\hsize}{-\algomargin}%
}
\makeatother

\def\T{{ \mathrm{\scriptscriptstyle T} }}

 \usepackage{geometry}
\begin{document}

\title{Quadruply robust estimation of marginal structural models in observational studies subject to covariate-driven observations}

\author[1]{Janie Coulombe}
\author[2]{Shu Yang}
\affil[1]{Department of Mathematics and Statistics, Université de Montréal, Montreal, Quebec, H3T 1J4 Canada}
\affil[2]{Department of Statistics, North Carolina State University, 2311 Stinson Drive, 5109 SAS Hall, Raleigh, United States }
 
\maketitle

\begin{abstract}
Electronic health records and other sources of observational data are increasingly used for drawing causal inferences. The estimation of a causal effect using these data not meant for research purposes is subject to confounding and irregular covariate-driven observation times affecting the inference. A doubly-weighted estimator accounting for these features has previously been proposed that relies on the correct specification of two nuisance models used for the weights. In this work, we propose a novel consistent quadruply robust estimator and demonstrate analytically and in large simulation studies that it is more flexible and more efficient than its only proposed alternative. It is further applied to data from the \textit{Add Health} study in the United States to estimate the causal effect of therapy counselling on alcohol consumption in American adolescents. 
\end{abstract}

 \section{Introduction}

The study of causes and effects is an important scientific topic and an essential component of learning healthcare systems aimed at improving public health (see e.g., \cite{krumholz2014big, dahabreh2014can}). When prescribing a treatment, physicians need to know about the causal effect of that treatment on the outcome to improve, instead of the mere association (or statistical dependence) between the treatment and outcome. Statistical methods with good properties must, therefore, be developed to estimate causal effects consistently with the least bias and variance possible. This manuscript proposes a novel, consistent and efficient estimator for the marginal causal effect of a treatment (exposure) on a longitudinal outcome using observational data.

Randomized controlled trials are the gold standard for drawing causal inferences. The randomization of participants to treatments ensures a balance in patient characteristics between treatment groups at study entry, allowing a fair comparison of clinical outcomes across treatment groups \citep{RN5910}. Often, randomized controlled trials also have clear protocols for the timing of patients’ visits (i.e., observation times) at which patient health status is measured. 

It is not always possible to conduct a randomized controlled trial designed specifically for answering a causal question, such that researchers often turn to observational data \citep{black1996we}. In this work, we focus on the particular features often met in observational data from electronic health records. Electronic health records data may contain information on patients’ demographic variables, diagnoses, health system usage, drug dispensations, and outcomes, making them highly useful to infer drug safety and effectiveness.
 
While electronic health records are increasingly available for analysis, they are not meant for research purposes. The treatments measured in electronic health records are not randomized to patients. Their observation rather depends on real-world prescription mechanisms, which depend on patients’ characteristics. This leads to spurious associations in the data called \textit{confounding} \citep{RN5693}. These data are also measured irregularly across patients. They are not collected under a controlled scheme. Instead, each patient follows their pattern in how they access care, also likely to depend on their characteristics \citep{lokku2020}. For instance, sicker patients tend to interact with the healthcare system more often than healthier patients. Statistically, this creates a long-term dependence structure between the outcome and the visit processes that can result in biased estimators of causal or associational parameters (see e.g., \cite{ lin2001semiparametric, lipsitz2002parameter, farewell2010marginal, pullenayegum2016longitudinal} and most recently \cite{coulombe2021estimating, coulombe, endogenous, yang2022semiparametric} in a context of causal inference). When aiming for a causal effect, this bias can be due to confounding by the visit process or, if the visit indicators act as colliders (i.e. are themselves affected by the treatment prescribed and the study outcome), to collider-stratification bias described in \cite{greenland2003quantifying}.

Under a set of causal assumptions, the causal marginal effect of a binary treatment on a longitudinal continuous outcome can generally be inferred by estimating the parameters of a marginal structural model fitted on the data from a pseudo-population that is free of confounding and other types of spurious associations, such as collider-stratification bias \citep{robins2000marginal}. Previous work has tackled this problem in settings with covariate-driven observation times and confounding, leading to the \textit{Flexible Inverse Probability of Treatment and Monitoring} weighted (FIPTM) estimator \citep{coulombe, endogenous}. However, that method suffers from two important problems. First, it relies on the correct specification of the treatment and outcome observation models as a function of patient characteristics. This implies both a correct specification of the variables to include in each model and their functional forms. When one or both models are not correctly specified, the FIPTM can be biased. Secondly, that estimator can be variable due to its inverse weights and could be made more efficient by using the geometry of influence functions and by augmenting the estimating equations \citep{robins1994estimation, RN5746}.   

To address the two issues raised above, we propose the first quadruply robust estimator for the causal marginal effect of a binary treatment on a longitudinal, continuous outcome, that accounts for confounding and irregular covariate-driven observation times of the outcome simultaneously. 
 
\section{Methods} \label{sec2}

\subsection{Notation}\label{not}

Our interest is in the causal marginal effect of a binary treatment or exposure (a choice between two possible options, say treatment with aspirin versus treatment with the absence of aspirin) on a longitudinal continuous outcome that is measured repeatedly. That effect could consist of a ``contrast'' causal effect between two different active drugs, or of the causal effect of an active drug when compared with a placebo. 

We assume working with a random sample of size $n$ from a larger population and denote by $i$ the patient index and by $t \in [0, \tau]$ the time, with $\tau$ a maximum follow-up time in the cohort. Let $A_i(t)$ represent the binary treatment taking values in $\left\{0,1\right\}$ and $Y_i(t)$ be the continuous, longitudinal study outcome for patient $i$ at time $t$, that is only observed at certain points in time. 

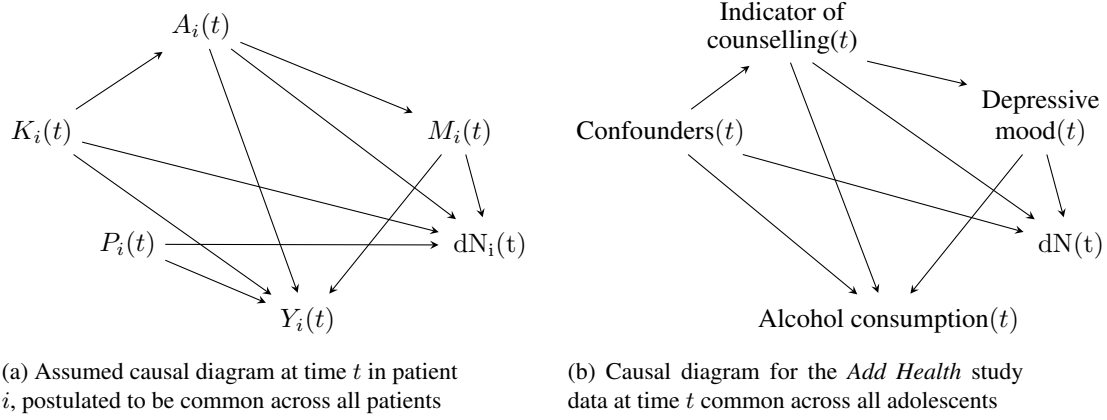
\begin{figure}[H]
\begin{subfigure}[b]{0.40\textwidth}
\begin{center}
\begin{tikzpicture}[%
->,
shorten >=2pt,
>=stealth,
node distance=1cm,
pil/.style={
->,
thick,
shorten =2pt,}
]
\node (1) at (-1.4,0.4) {$ A_i(t) $};
\node (2) at (2,-1) {$M_i(t)$};
\node(3) at (0,-3.5) {$Y_i(t)$};
\node (4) at (2.4,-2.5) {$\de N_i(t)$};
\node(5) at (-3.5,-1){$K_i(t)$};
\node(88) at (-2.37, -2.49) {$P_i(t)$};

\draw[->] (1) to  (3);
\draw[->] (2) to  (3);
 \draw[->] (1) to (2);
\draw[->](5) to  (1);
\draw[->](5) to (3);
\draw[->,black] (1) to (4);
\draw[->,black] (2) to  (4);
 \draw[->] (88) to  (3);
  \draw[->] (88) to  (4);
 \draw[->](5) to (4);
\end{tikzpicture}
\caption{Assumed causal diagram at time $t$ in patient $i$, postulated to be common across all patients} \label{panelA}
\end{center}
\end{subfigure}
\hspace{.5in} 
\begin{subfigure}[b]{0.40\textwidth}
\begin{center}
\begin{tikzpicture}[%
->,
shorten >=2pt,
>=stealth,
node distance=1cm,
pil/.style={
->,
thick,
shorten =2pt,}
]
\node (1) at (-1.4,0.6) {Indicator of};
\node (11) at (-1.4,0.2) {counselling($t$)};
\node(2b) at (2,-0.6){Depressive};
\node (2) at (2,-1) {mood$(t)$};
\node(3) at (0,-3.5) {Alcohol consumption$(t)$};
\node (4) at (2.4,-2.5) {$\de N(t)$};
\node(5) at (-3,-1){Confounders$(t)$};

\draw[->] (11) to  (3);
\draw[->] (2) to  (3);
 \draw[->] (11) to (2b);
\draw[->](5) to  (11);
\draw[->](5) to (3);
\draw[->,black] (11) to (4);
\draw[->,black] (2) to  (4);
\draw[->,black] (5) to (4);
\end{tikzpicture}
\caption{Causal diagram for the \textit{Add Health} study data at time $t$ common across all adolescents} \label{panelB}
\end{center}
\end{subfigure}
 
\caption{Causal diagram illustrating the assumed data generating mechanism}\label{fig1b}

\end{figure}

The type of data-generating mechanism we focus on is presented in the left panel of Fig. \ref{fig1b}. The proposed approach is not specific to that data-generating mechanism and the novel estimator could be tailored according to various data-generating mechanisms. 
In Fig. \ref{fig1b}, variables $K_i(t)$ are confounders that create (open) backdoor paths from the treatment $A_i(t)$ to the outcome $Y_i(t)$. Associations that are not due to the causal effect of the treatment on the outcome can pass through them \citep{pearl2009causality}. The set $M_i(t)$ denotes a set of mediators for the treatment effect on the outcome. The set $P_i(t)$ contains pure predictors of the outcome that potentially also affect the observation of a patient outcome $Y_i(t)$.

We use a counting process to model observation times. Only the outcome process is assumed to be measured sporadically (one may think of weight or systolic blood pressure measured irregularly according to an observation scheme depending on patient characteristics such as a change in medication or smoking status). All the other variables necessary to the estimation of the marginal causal effect of treatment are assumed to be available at all times during follow-up, for each patient. While this may seem like a strong assumption,  the drugs and comorbidities are typically recorded in electronic health records data anytime there is a new diagnosis or a new prescription is made or dispensed. Let $N_i(t)$ be the total number of observation times of the outcome $Y_i(t)$ between times 0 and $t$ for individual $i$. The indicator $\de N_i(t)$ equals to 1 when there is a jump in the process at time $t$, i.e., the observation of the outcome $Y_i(t)$, and 0 otherwise. We denote by $T_{i,1},..., T_{i,Q_i}$ the observation times of the outcome, with $Q_i$ the total number of observation times of individual $i$. The set $V_i(t)$ includes all the variables causing observation times (i.e., causing $\de N_i(t)$ in Fig. \ref{fig1b}) and \textit{also} including all the confounders of the treatment-outcome relationship. We must include $K_i(t)$ in the set $V_i(t)$ for the proposed methodology to be consistent. In Fig. \ref{fig1b}, the set $V_i(t)$ contains the treatment, the mediator(s) of the treatment effect on the outcome, the confounder(s), and the pure predictor(s).

Patients are allowed to have different follow-up times $C_i$, assumed to be non-informative conditional on the outcome model design matrix, an assumption denoted by $Y_i(t) \perp C_i \mid A_i(t), V_i(t)$. Finally, $\xi_i(t)=\bone\{ C_i>t \}$ is an indicator that patient $i$ is still in the study at time $t$, meaning they are not lost to follow-up. 

 \subsection{Causal Estimand}\label{cauestimand}

The potential outcome framework  \citep{neyman1923application, rubin1976inference} is used to define our estimand. Denote by $Y^1_i(t)$ the potential outcome of individual $i$ at time $t$ if they received treatment option $1$, and $Y^0_i(t)$ for treatment $0$. The causal marginal effect of a binary treatment on a continuous outcome is defined as  $\beta_1=\E\left[ Y^1_i(t) - Y^0_i(t) \right]$. The treatment and the outcome are allowed to vary in time, but our interest lies in a cross-sectional effect $\beta_1$ that does not vary in time.

Suppose a certain time discretization for which there can be only one jump in the counting process $N(t)$. For instance, suppose visits at the doctor's office can occur daily such that the time granularity is the day. If we had access to all potential outcomes under both treatments and at each time $t$ (for the time granularity chosen above) for a random sample of participants of size $n$, then we could estimate $\beta_1$ straightforwardly using sample means. This is impossible in practice and sometimes referred to as the fundamental problem of causal inference \citep{holland1986statistics}. 

On the other hand, by conducting a randomized controlled trial and randomly allocating patients to one of the two treatment options, and observing patients at prespecified visit times, there is no reason why patients allocated to, e.g., treatment $1$, would differ than the others before receiving the treatment. One could then use the marginal structural model to estimate $\beta_1$: 
 \begin{equation}
      \E\left[ Y^a_i(t)\right] = \E\left[ Y^a_i(t)  \mid A_i(t)=a\right] 
      = \beta_0 + \beta_1 a. \label{hg}
 \end{equation}
Under this model, $\E\left[ Y^1_i(t) \right]= \beta_0 + \beta_1 $ and $\E\left[ Y^0_i(t)\right]= \beta_0$, and the causal contrast of interest is the parameter $\beta_1$. An outcome consistency assumption (discussed in $\mathsection$\ref{causalass}) is required to estimate $\beta_1$ with the marginal structural model above, but it generally does not require any adjustment for confounding under random treatment allocation.

In observational data from electronic health records, unfortunately, data are affected by a confounding mechanism, whereby we observe the potential outcomes $Y^1_i(t)$ in those who had greater chances of being treated with $A_i(t)=1$, and the potential outcomes $Y^0_i(t)$ in those who had greater chances of being treated with $A_i(t)=0$ (as a consequence, $\E\left[ Y^a_i(t)  \mid A_i(t)=a \right] \neq \E\left[ Y^a_i(t)\right]$ in general). In addition to that confounding, the potential outcomes for an individual $i$ are only observed at times $t$ when $\de N_i(t)=1$, which may also depend on patient characteristics. Therefore, we do not have access to all potential outcomes and require a set of causal assumptions to equate the estimand, which depends on $Y_i^a(t)$, to equations depending on $Y_i(t)$. 

\subsection{Causal Assumptions}\label{causalass}
Five causal assumptions are required for consistent estimation of the causal marginal effect of treatment (Table \ref{tabass}). Other (non-causal) assumptions on the correct specification of the nuisance models are also required and discussed in $\mathsection$\ref{st}.  

First, it requires outcome consistency, which implies that $Y_i(t)=Y^1_i(t)$ if patient $i$ received treatment 1 at time $t$ and  $Y_i(t)=Y^0_i(t)$ otherwise and means that the treatment is well defined such that the observed outcome effectively corresponds to one of the two potential outcomes. An example of consistency violation is one in which a potential outcome is defined as the outcome under the treatment \textit{physical exercise 5 times a week} and in which the actual treatment received corresponds to \textit{having done exercise 3 times a week}. The outcome observed under that regimen would not correspond to the defined potential outcome.

We assume positivity of treatment, meaning that anyone should have a chance of receiving any of the two treatment options, and positivity of observation, such that they had a chance to have their outcome observed at any time given their characteristics. For instance, scenarios in which some patient characteristics used as predictors in the treatment model are only represented in one of the two treatment groups would violate the positivity of treatment assumption, and similarly for the observed and non-observed groups.  

Finally, we assume conditional exchangeability, with includes the assumptions of 1) no unmeasured confounder, i.e., all confounders $K_i(t)$ of the relationship between the treatment and the outcome are available in the analysis; and 2) conditional independence of the observation indicators, i.e., adjusting for $V_i(t)$ makes the observation indicator independent of other variables in the analysis. In our setting, $K_i(t) \subset V_i(t)$ and exchangeability is conditional on $V_i(t)$. 

These five assumptions allow the use of so-called \textit{G-methods} \cite[Ch.\ 23, p. 553]{naimi2017introduction,fitz} for consistent estimation of causal effects. The methods discussed next can be thought of as being part of a larger \textit{G-methods} framework.

\begin{table}[H]
\caption{Causal assumptions required for the proposed estimator to be consistent}

\centering
\resizebox{\textwidth}{!}{%
\begin{tabular}{|l| c|} 
\hline 
   Assumption & Definition \tabularnewline
\hline 
\hline 
  Outcome consistency &  $Y_i(t) = A_i(t) Y_i^1(t) + \left\{1-A_i(t)\right\} Y_i^0(t)$\tabularnewline
\hline 
  Positivity of treatment & $0 < \pr \{ A_i(t)\mid K_i(t) \}< 1$ \tabularnewline
\hline  
  Positivity of observation & $0 < \E[\de N_i(t)\mid V_i(t)]< 1$\tabularnewline
\hline
  No unmeasured confounder & $\left\{ Y_i^0(t), Y_i^1(t)\right\}\perp A_i(t)  \mid K_i(t) $ \tabularnewline
\hline
  Conditional exchangeability & $\left\{ Y_i^0(t), Y_i^1(t)\right\}\perp A_i(t)  \mid K_i(t) $ and $\{ A_i(t), Y^0_i(t), Y^1_i(t)\} \bot  \de N_i(t)   \mid  V_i(t) $ \tabularnewline
\hline
\end{tabular} }\label{tabass}
\end{table}

 \subsection{Novel Estimator}\label{st}
 
 The conditional exchangeability can be recovered by breaking the spurious associations due to the treatment and observation mechanisms via inverse weights (marginal approach), by conditioning on the sets $A_i(t), K_i(t), V_i(t)$ in a regression model for the outcome and using methods such as g-computation \citep{RN5748} (standardization approach), or by using both approaches simultaneously to make it more robust to models misspecification, which is our proposal.

Using the marginal approach corresponds to using the FIPTM estimator proposed in \cite{coulombe}.  It consists of a doubly-weighted least squares estimator that incorporates inverse probability of treatment weights  \citep{RN5605,RN5668,RN5606}  and inverse intensity of visit weights \citep{lin2004analysis}. The inverse probability of treatment weights are functions of the confounders $K_i(t)$ and the IIV weights are functions of the visit predictors $V_i(t)$. The estimator is consistent for $\beta_1$ when both weights are correctly specified. A parametric model can be used to model the treatment, and the inverse probability of treatment weights be obtained as follows:
\begin{align}
 \bone\{A_i(t)=a\}/\pr\{A_{i}(t)=a\mid K_{i}(t);\psi\} \label{iptt}
\end{align}
where $\pr\{A_{i}(t)=1\mid K_{i}(t);\psi\}$, the propensity score, is the probability of receiving the treatment $1$ as a function of predictors $K_{i}(t)$ and parameters $\psi$ \citep{RN5668}. A logistic regression can be used to compute an estimated propensity score. The inverse intensity of visit weights, on the other hand, can be obtained from the proportional rate model:
\begin{align}
\E[ \de N_{i}(t)\mid V_{i}(t); \gamma]=\xi_{i}(t)\exp\{\gamma^{\T}V_{i}(t)\}\lambda_{0}(t)\de t. \label{eqq}
\end{align}
The baseline rate of observation $\lambda_{0}(t)$ in (\ref{eqq}) consists of the visit rate when all variables $V_i(t)$ are set to their reference level. With the FIPTM estimator, the baseline rate can be dropped from the inverse intensity of visit weights without affecting the causal marginal effect of treatment estimate since removing it would still make the weights in (\ref{eqq2}) proportional to the intensity of being observed as a function of $V_{i}(t)$. The inverse intensity of visit weights can also be stabilized, in which case the baseline rate cancels automatically in the weights and need not be estimated \citep{buuvzkova2009semiparametric}. This leads to the following intensity of visit weights (we take the inverse in the estimating equations), from which $\gamma$ parameters can be estimated using the \cite{andersen1982cox} model:
\begin{align}
 \E [\de N_i(t) \mid V_i(t); \gamma ]= \exp\{\gamma^{\T}V_{i}(t)\}. \label{eqq2}
\end{align}
 Then, the FIPTM estimator solves the following equation: 
\begin{align}
\E_{n}\left[\int_{0}^{\tau}\frac{\frac{\bone\{A_{i}(t)=a\}}{\pr\{A_{i}(t)=a\mid K_{i}(t);\hat{\psi}\}}Y_{i}(t)-\zeta_{i}(t;\beta_{a})}{\E \{ \de N_i(t) \mid V_i(t); \hat{\gamma} \}}\de N_{i}(t)\right]=0, \label{eqq6}
\end{align}

\noindent where $\zeta_{i}(t;\beta_{a})=\beta_0 + \beta_1 a$ is the structural model from equation (\ref{hg}) and $\E_n$ stands for the empirical mean. A drawback of the doubly-weighted estimator is that it requires both the treatment and the observation models to be correctly specified. This is not easy in practice. 

We propose the augmented AAIIW estimator (which acronym stands for \textit{doubly augmented and doubly inverse weighted}) that is more flexible and allows two out of four different models (see Table \ref{tab1}) to be misspecified while the estimator remains consistent. The estimator uses the theory introduced in \cite{robins1994estimation}, very well laid out in \cite{funk2011doubly}, \cite{tsiatis2007comment} and \cite{cao2009improving}, and further related to model-assisted estimation from the survey sampling field (see e.g., the discussion in \cite{chambers1998discussion}). Using the model-assisted estimation approach to justify the construction of the novel estimator may be more intuitive to the reader than using the semiparametric theory of influence functions (see e.g., \cite{jiang2022multiply} for related discussions), so we briefly discuss that framework. \textit{Grosso modo}, the estimating equations of the FIPTM estimator are to be transformed twice following the model-assisted estimation approach. Based on their work in \cite{robins1994estimation}, \cite{chambers1998discussion} introduce the equation 
\begin{align*}
N \hat{T}_{diff}(\mu) = \sum_{i=1}^N A_i \mu(X_i) + \sum_{i:A_i=a} A_i \left\{ Y_i - \mu(X_i)\right\}/ \pi_i 
\end{align*}
and connect it to model-assisted estimation, where $ \hat{T}_{diff}(\mu)$ is a designed-based standard difference estimator for the parameter $\E[A_i Y_i]$ of interest, and $\mu(X)$ is a function of $X$. Theorem 1 in \cite{chambers1998discussion} implies that the class of such estimators $\hat{T}_{diff}(\mu)$ contains all the semiparametric estimators and that the asymptotic variance of the estimator evaluated at $\mu_{eff}(x)= \E[Y_i\mid X_i=x]$ in their notation leads to the smallest variance possible. We use that construction twice, starting first with $A_i = \bone\{ A_i(t) = a\}$, $\pi_i = \pr\{A_i(t)=a\mid K_i(t) \}$ and $``\mu(X_i)"= \E \left[ Y_i(t) \mid A_i(t)=a, K_i(t)   \right] $ using their notation. Once this projection is obtained, we use the strategy a second time, with the ``$Y_i$" now corresponding to the previous expression obtained, and with $``A_i" =\de N_i(t)$, $``\pi_i" = \pr\{\de N_i(t)=1\mid V_i(t) \}$ and a novel $``\mu(X_i)"$ function corresponding to the expectation of the previous term before augmentation (in equation (\ref{eg1}) below, this expectation corresponds to $\E[\eta_i(t) \mid A_i(t)=a, K_i(t), V_i(t)   ]$). By construction, the novel proposed estimator is the most efficient among its class of semiparametric estimators, which also includes the FIPTM estimator.

The correspondence between the model-based estimation approach and the geometry of influence functions is in that the novel estimator's influence function corresponds to sequential projections of the FIPTM's influence function onto spaces orthogonal to the residuals from the weight models and orthogonal to projections due to the outcome mean models onto the spaces of $K_i(t)$ and $V_i(t)$. The novel estimator is obtained by solving the following estimating equations, which are augmented versions of the equations in \ref{eqq6}:
\begin{eqnarray}
 &  & \E_n \left[\int_{0}^{\tau}\frac{ \eta_i(t) }{\E \{ \de N_i(t) \mid V_i(t); \hat{\gamma} \}}\de N_{i}(t)\right]   -  \E_n \left[ \int_{0}^{\tau}\frac{\de M_{i}(t)   \E \{ \eta_i(t)\mid   A_i(t)=a, K_i(t), V_i(t)   \} }{\E \{ \de N_i(t) \mid V_i(t); \hat{\gamma} \} }  \right]  \nonumber \\ 
 & &  =0,  \label{eg1}
\end{eqnarray}
where the nuisance terms are estimated using parametric models, with
\begin{eqnarray*}
    \eta_i(t) = &\frac{\bone\{ A_{i}(t)=a\}}{\pr\{A_{i}(t)=a\mid K_{i}(t);\hat{\psi}\}}Y_{i}(t){\color{black}-\frac{\bone\{A_{i}(t)=a\}-\pr\{A_{i}(t)=a\mid  K_{i}(t);\hat{\psi}\}}{\pr\{A_{i}(t)=a\mid  K_{i}(t);\hat{\psi}\}}\mu_{a}\{ K_{i}(t);\hat{\alpha}_{K}\}}-\zeta_{i}(t;\beta_{a}),
    \end{eqnarray*}
with $\de M_{i}(t)=\de N_{i}(t)-\xi_{i}(t)\exp\{\hat{\gamma}^{\T}V_{i}(t)\}\hat{\lambda}_{0}(t)\de t$ the martingale residual for the observation process. The conditional outcome mean models in the augmentated terms are $\mu_a\left\{K_i(t); \alpha_K\right\}=\E[Y_i(t)\mid A_i(t)=a, K_i(t); \alpha_K]$ and  $\mu_a\left\{V_i(t); \alpha_V\right\}=\E[Y_i(t)\mid A_i(t)=a, V_i(t) ; \alpha_V]$. The latter model arises when taking the expectation $\E \left[ Y_i(t) \mid A_i(t)=a, V_i(t)\right]$ in the term $\E \{ \eta_i(t)\mid A_i(t)=a, K_i(t), V_i(t) \}$ in equation \ref{eg1} (more details are found in Supplementary Material A and B discussed later). For the novel estimator, the baseline rate  $\lambda_{0}(t)$ in (\ref{eqq}) must be estimated before calculating the inverse intensity of visit weights. The inverse intensity of visit weights in the equations for the AAIIW are the inverse of $\E [\de N_i(t) \mid V_i(t); \hat{\gamma} ]=\hat{\lambda}_0(t)\exp\{\hat{\gamma}^{\T}V_{i}(t)\}$. We use the following Breslow's estimator:

 \begin{align*}
\hat{\lambda}_0(t) = \frac{  \sum_{i=1}^n   \bone \{ \de N_i(t)=1 \}   }{  \sum_{i=1}^n  \bone \{ \de N_i(t)=1\} \exp \left\{ \hat{\gamma}^{T} V_i(t) \right\}  },
 \end{align*}

\citep{cox1972regression}.  Table \ref{tab1} shows the possible combinations of correctly specified models leading to a consistent AAIIW estimator. For the estimator to be unbiased, at least one of the two models related to confounders (either the treatment or the outcome mean model conditional on the confounders) and at least one of the two models related to the observation predictors (either the observation or the outcome mean model conditional on the observation predictors) must be correctly specified. Correct specification for a nuisance model requires that the corresponding data-generating mechanism can be modelled parametrically and that there exists a true set of parameters leading to the actual data-generating mechanism that we can estimate consistently (we denote the true sets by $\psi_0$, $\gamma_0$, $\alpha_{K0}$ and $\alpha_{V0}$ for the treatment, observation, and two conditional mean outcome models, respectively). For instance, for the treatment model, it means that the data generating mechanism $\pr\{A_{i}(t)=a\mid  K_{i}(t) \}=\pr\{A_{i}(t)=a\mid  K_{i}(t); \psi_0\}$, that we can model this data generating mechanism using the correct functional formats for covariates $ K_{i}(t)$ in the model, and that estimators $\hat{\psi}$ converge in probability to the true parameters $\psi_0$.

Naturally, the AAIIW estimator is unbiased under specific combinations of two correctly specified models (Table \ref{tab1}, see the proof of multiple robustness in Supplementary Material A) \textit{but also} when three out of the four models in Table \ref{tab1} are correctly specified, or when all models are correctly specified. Thus, using the same nuisance models as those used in the FIPTM estimator, if they are correctly specified, leads to an unbiased estimator, but the advantage of the AAIIW estimator is that it also has several other opportunities to be unbiased. 
 \begin{table}
\caption{Quadruple robustness of AAIIW: AAIIW is consistent under Scenarios
(a)--(d): $\checked$ means correctly specified and $X$ means no
requirement}
\centering
\resizebox{\textwidth}{!}{%
\begin{tabular}{|c|c|c|c|c|} 
\hline 
Scenario & $\pr\{A_{i}(t)=a\mid K_{i}(t);\psi\}$ & $\mu_{a}\{K_{i}(t);\alpha_{K}\}$ & $\E\left[\de N_i(t)=1\mid V_i(t);\gamma\right]$ & $\mu_{a}\{V_{i}(t);\alpha_{V}\}$\tabularnewline
\hline 
\hline 
(a) & $\checked$ & $X$ & $\checked$ & $X$\tabularnewline
\hline 
(b) & $X$ & $\checked$ & $X$ & $\checked$\tabularnewline
\hline 
(c) & $X$ & $\checked$ & $\checked$ & $X$\tabularnewline
\hline 
(d) & $\checked$ & $X$ & $X$ & $\checked$\tabularnewline
\hline 
\end{tabular}} \label{tab1}
\end{table}

In practice, the AAIIW estimator can be computed by first estimating the parameters from all nuisance models (using a logistic regression, a proportional rate model with e.g., \texttt{coxph} in R, and two linear models for the outcome means conditional on $K_i(t)$ or $V_i(t)$). The estimates can be plugged into the estimating equations of the AAIIW. A root solver such as \texttt{uniroot} in R can be used then to estimate $\beta_0$ and that estimate be further plugged into the second estimating equation, which is solved for $\beta_1$.
\subsection{Efficiency and Asymptotic Properties}

 The asymptotic variance of the AAIIW estimator can be derived using theory on two-step estimators \citep{newey1994handbook} or deriving the variance of its influence function \citep{RN5746}. We use the latter approach. Derivations are shown in Supplementary Material B, in which we also show that the AAIIW estimator is more efficient than the FIPTM estimator when all nuisance models are correctly specified for both estimators. We find that under correct nuisance models specification, the FIPTM asymptotic variance equals to
\begin{eqnarray*}
\sigma^2_{FIPTM}= \E \left[  \frac{   \{Y^1_i(t) - \mu_1\}^2  }{\rho\{ V_i(t) \} e_1  \{ K_i(t) \}} \right]         + \E \left[  \frac{  \{Y^0_i(t) - \mu_0\}^2  }{\rho\{ V_i(t) \} e_0 \{ K_i(t) \} } \right],
 \end{eqnarray*}
 where $\mu_a = \E[ Y^a_i(t)]$, $\rho\{ V_i(t) \}= \E [\de N_i(t)\mid V_i(t); \gamma_0 ] $, and $e_a \{ K_i(t) \}=\pr\{ A_i(t)=a\mid K_i(t); \psi_0 \}$. The augmented AAIIW estimator is more efficient, with asymptotic variance
\begin{eqnarray*}
\sigma^2_{AAIIW} = \sigma^2_{FIPTM} - \mu_1^2 \E \left[   \frac{1+e_1\{ K_i(t) \}}{e_1\{ K_i(t) \}}  \right]- \mu_0^2 \E \left[   \frac{1+e_0\{ K_i(t) \}}{e_0\{ K_i(t) \}}  \right] .
 \end{eqnarray*}
Furthermore, the theory of influence function can be used to show the asymptotic normality of the estimator. The AAIIW estimator is consistent and converges to the true causal effect of the binary treatment, so its limiting distribution is normally distributed around $\beta_1$.
   
\section{Simulation study}\label{sec3}
 
\subsection{Comparators}

 We compared four different estimators in large simulation studies: an ordinary least squares estimator that does not account for confounding or informative observation (OLS); an inverse probability of treatment-weighted estimator that accounts for the confounding process with a propensity score that is either correctly specified (IPT$_c$) or not correct (IPT$_{nc}$); a doubly-weighted estimator that corresponds to the FIPTM estimator from \cite{coulombe} with either both the inverse intensity of visit and inverse probability of treatment weights correctly specified (DW$_{c}$), only the inverse probability of treatment weights correctly specified (DW$_{iptc}$), only the inverse intensity of visit weights correctly specified (DW$_{iivc}$), or both weights misspecified (DW$_{nc}$); and the novel AAIIW estimator with either all four nuisance models correctly specified (AAIIW$_c$), both weight models correctly specified and both conditional outcome models misspecified (AAIIW$_{s.a}$, with s.a referring to \textit{scenario a} in Table \ref{tab1}), both conditional outcome models correctly specified and both weights misspecified (AAIIW$_{s.b}$), the inverse intensity of visit weights and the outcome model conditional on confounders being the only correctly specified models (AAIIW$_{s.c}$), and the inverse probability of treatment weights and the outcome model conditional on observation predictors being the only correctly specified models (AAIIW$_{s.d}$). 
 
 The data-generating mechanism was strongly inspired by similar simulation studies presented in \cite{buuvzkova2009semiparametric, coulombe, endogenous} and is described in much more detail in Supplementary Material C. The data-generating mechanism included a set of confounders at baseline repeated through follow-up, a time-varying binary treatment, a set of observation predictors that varied in time, and irregular observation of the outcome. The main results for 1000 simulations using a nonhomogeneous Poisson rate to simulate the observation times of the outcome and a sample of size 1000 are presented in the following section. In another set of simulations, we replaced the nonhomogeneous Poisson rate with a nonhomogeneous Bernoulli probability for the observation indicator and used a logistic regression instead of the Andersen and Gill model to fit the probability of observation at each time point. These results are presented in Supplementary Material D (Suppl. Fig. 2 and 3), along with the results under a sample of size 250 instead of 1000 (Suppl. Fig. 1) and all Monte Carlo biases and mean square errors (Suppl. Table 1). In both settings using either the Poisson rate of the Bernoulli probability, we tested four different sets of $\gamma$ parameters in the observation model, including one set of zeros (which we call ``set 1" in the results) corresponding to uninformative observation.

\subsection{Results}\label{res}

\begin{figure} 
\begin{center}
\includegraphics[width= 0.9\textwidth]{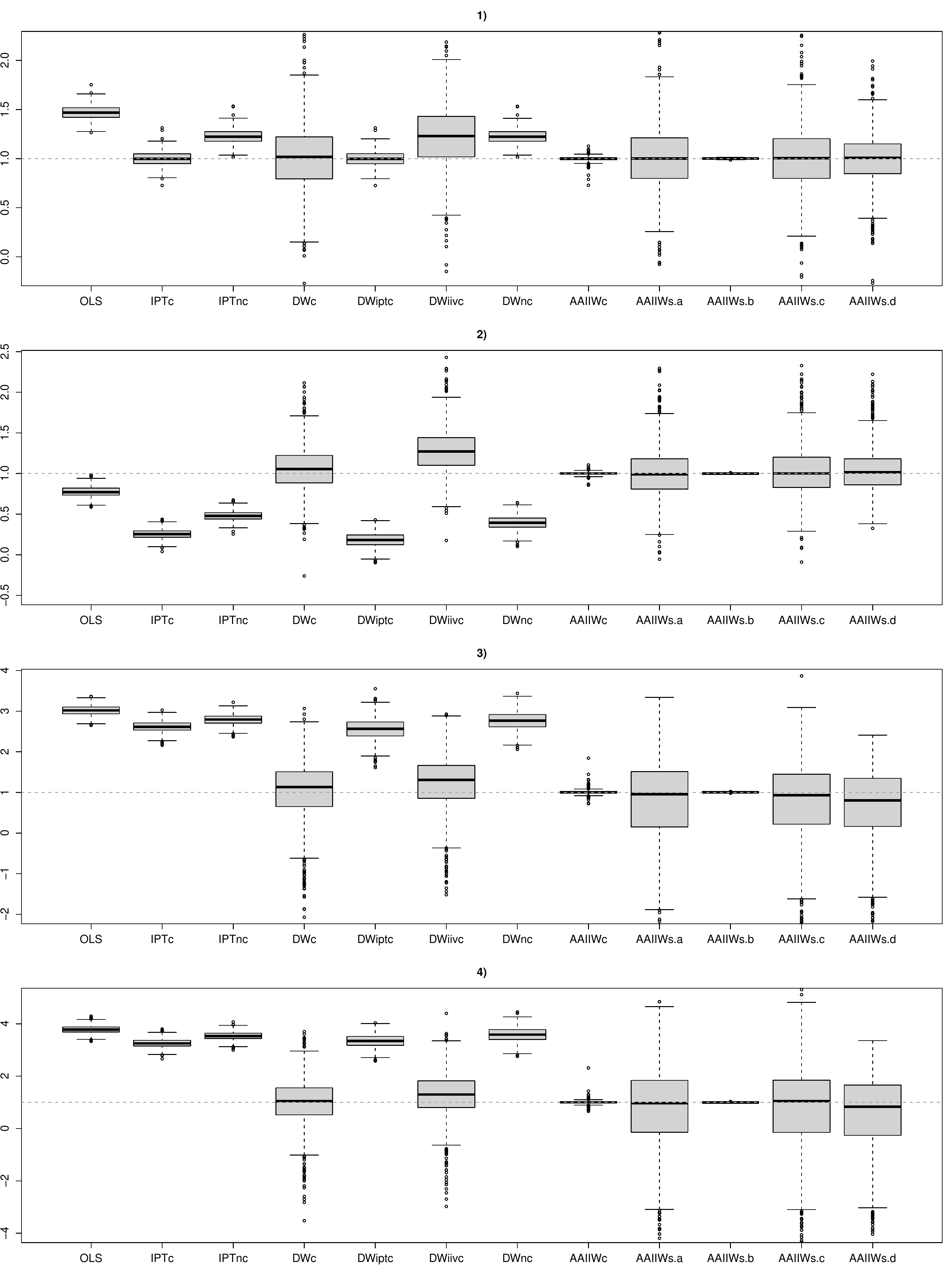}   
 \caption{Results of the simulation studies with a sample size of 1000 using a nonhomogeneous Poisson rate to simulate the observation indicators and the Andersen and Gill model with Breslow estimator to estimate the inverse intensity of visit weights. Each boxplot represents the distribution of 1000 estimates for the corresponding estimator. The dashed line represents the gold standard, i.e., the true value for the marginal effect of exposure that equals to 1. Different strengths of the visit process on covariates are represented with scenarios 1) $\gamma=(0, 0, 0, 0, 0, -5)$ (i.e., no bias due to the visit process expected); 2) $\gamma=(0.5, 0.3, -0.5, -2, 0, -3)$; 3) $\gamma=(0.5, -0.5, -0.2, -1, 1, -3)$; and 4) $\gamma=(-1, -0.8, 0.1, 0.3, -1, -3)$.}\label{fig2}
\end{center}
\end{figure}

The distributions of 1000 estimates obtained with each estimator using a sample of size 1000 patients are presented in Fig. \ref{fig2}. 

The results are as expected. First, the ordinary least squares estimator is strongly biased in all $\gamma$ parameter settings 1) to 4). In scenario 1) in which we expected no bias due to the visit process, the inverse probability of treatment-weighted estimator IPTc is empirically unbiased and the inverse probability of treatment-weighted estimator using a wrong treatment model, IPTnc, exhibits bias. In scenarios 2) to 4), both inverse probability of treatment-weighted estimators are biased since they do not account properly for the outcome observation process.

The doubly-weighted estimator DWc is consistently unbiased as it accounts properly for both types of bias. When the visit process is uninformative, in scenario 1) for the observation process, it is also unbiased even when the inverse intensity of visit weights are not correctly specified, as long as the inverse probability of treatment weights are correctly specified. In scenarios 2) to 4), the doubly-weighted estimator is only unbiased in settings in which its two weight models are correctly specified.

The multiply robust AAIIW estimator is empirically unbiased in all scenarios 1) to 4) for the observation process, whenever using one of the four combinations of correctly specified models shown in Table \ref{tab1} or when all four nuisance models are correctly specified. It exhibits particularly small variance when the two conditional outcome mean models are correctly specified  (scenario b from Table \ref{tab1}) or, as expected when all four models are correctly specified.

Results for the second set of simulations using the Bernoulli probability to simulate the observations, and those for a sample of size 250 are presented in Supplementary Material D. As expected, the estimators are more variable when using a sample size of 250, although the same patterns in the comparison of estimators are observed (Suppl. Fig. 1). Similar results are observed when using the Bernoulli probability instead of the Poisson rate for the simulation of observation indicators (Suppl. Fig. 2 and 3). The simulations using the Bernoulli probability did not require the use of Breslow's estimator for the baseline rate, which would partly explain the smaller variances observed overall (e.g., compare Fig. 1 and Suppl. Fig. 2).

 \section{Motivating Example} \label{appli}
 
 We applied the proposed AAIIW estimator and different more naive comparators to data from the \textit{Add Health} study in the United State \citep{harris1, harris2, harris3}. It consists of a longitudinal study with multiple waves. The study started in 1994 when a pool of adolescents representative from the United States was selected. These adolescents grew up to become adults during the study. At each wave, they were asked to fill out questionnaires.

 We have access to data from the first four waves of the \textit{Add Health} study, corresponding to the years 1994-1995, 1996, 2001-2002 and 2008-2009 respectively. 
 Although these data do not consist of electronic health records, and that adolescents in the \textit{Add Health} study were observed at pre-specified observation times, they tended not to be observed at each wave for all variables measured in the study, and one could consider that the four waves represent four consecutive time points when there could be or no a visit or an observation time, just like in medical records. Various types of information such as demographics, health status, nutrition, family dynamics, sexual activity and substance use were collected for the study. Some questions varied across the four waves and we focused on a causal research question for which we have data at all four waves. Our goal was to estimate the marginal causal effect of counselling (psychotherapy) on alcohol consumption. We think that the effect of counselling on alcohol consumption is mediated by the depressive mood of adolescents and that their mood can be affected by counselling and may in its turn affect alcohol consumption (see the assumed data generating mechanism in Fig. \ref{panelB}). Two important challenges we wished to consider in the analysis are the irregular observation of the outcome and, because the study is observational, the potential confounding of the psychotherapy-alcohol consumption relationship. 

We selected potential confounders for that relationship, which included the teen's age, sex, socioeconomic status, weight in pounds, and whether they smoked at least once in the previous month. The socioeconomic status was computed by summing two variables that we transformed, namely the parents' total income in 1994 before taxes and one of the parents' education, usually that of the resident mother \citep{harris1}. The parent's total income was transformed into quintiles (1 to 5 with 5 being the highest). One of the parents' education was categorized in 5 levels corresponding to \textit{1- 8th grade or less or never went to school}, \textit{2-more than 8th grade but did not graduate high school}, \textit{3-went to a business, trade or vocational school instead of high school, high school graduate, or completed a general educational development program}, \textit{4--went to a business, trade or vocational school after high school, went to college but did not graduate}, or \textit{5-graduated from college or university, or professional training beyond a 4-year college or university training}. Socioeconomic status was defined as the sum of the two transformed 5-category variables.

\textcolor{black}{The analysis dataset contained several missing values. Unless we had enough information in the dataset to replace missing values in the variables age and sex (e.g., if age was measured at a previous wave and it could be used for extrapolation), we used multiple imputations by chained equations  \citep{rubin1988overview,azur2011multiple} to impute missing values in these variables as well as in variables socioeconomic status, smoking status, weight, depressive mood, and the exposure to counselling. We used these variables as predictors each time to impute each variable using fully conditional specification. The alcohol outcome was kept as missing when it was not measured. } 

 The outcome was defined using the question \textit{Think of all the times you had a drink during the past 12 months. How many drinks did you usually have each time?}. It consisted of a self-assessed number of drinks the adolescent would consume, on average, each time they consumed alcohol. It ranged from 0 to 90. In this application, the outcome tended to be assessed at each of the four waves (i.e., not irregularly and with most data not being missing in the outcome). To assess the advantage of our approach, we simulated missingness in the outcome and assessed the different estimators in that setting, knowing the true underlying missingness mechanism.
  
Assuming that all potential confounders as well as the mediator (depressive mood) and the exposure (counselling) affect the chance of observing the alcohol consumption outcome, the outcome observation (i.e., the opposite of missingness) was simulated using the following model across the four waves:
\begin{align*}
 &\E[\de N_i(t)\mid \text{age, sex, counselling, depressive mood, socioeconomic status}]
 = \text{expit}\{ 18 -  \\ & 0.3\hspace{0.1cm}\text{age}(t)+0.8\hspace{0.1cm}\text{sex} + 1.8 \hspace{0.1cm}\text{counselling}(t) -3 \hspace{0.1cm} \text{depressive mood} (t) -1.3 \hspace{0.1cm} \text{socioeconomic status} \}
 \end{align*}
for $t \in 1, 2, 3, 4$, where expit$(\cdot)=\exp(\cdot)/\left\{1+\exp(\cdot)\right\}$. 
   
 We conducted the analysis and fit a propensity score model as a function of age, sex, weight, socioeconomic status and smoking. We fit two different proportional rate models for the observation of the outcome, one correctly specified (as a function of age, sex, counselling, depressive mood, and socioeconomic status) and one that was not correctly specified (as a function of the sinus of age and the depressive mood, therefore not including the right format for the age variable and missing some important variables in the model). The estimators compared in the application are a standard ordinary least squares estimator (not adjusted for confounding nor for the observation process), an inverse intensity of visit-weighted estimator that does not account for confounding but does account for the observation process (we tested the two sets of the inverse intensity of visit weights), a doubly-weighted estimator corresponding to the FIPTM estimator (incorporating the inverse probability of treatment weights based on our assumptions on the potential confounders, and inverse intensity of visit weights - we tested the two sets of the inverse intensity of visit weights here again), and the AAIIW estimator in which we incorporated the inverse probability of treatment weights and the two different sets of the inverse intensity of visit weights, one at a time.
    
We found an important difference in the proportion of males and females across both exposure groups, with females reporting higher rates of counselling, $14\%$ more smokers in the counselling group than the other, and a greater depressive mood in those receiving counselling (Supplementary Material E, Suppl. Table 2). These variables were, therefore, considered potential confounders in our analysis, except for the mediator depressive mood that should not be conditioned upon in the confounding set. After inverse probability of treatment-weighting, the two exposure groups are similar with respect to all potential confounders (Suppl. Table 2).

In the outcome observation model, we found modest differences in the depressive mood and counselling rates between those for whom the alcohol consumption was observed and the others (Supplementary Material E, Suppl. Table 3). There was also a slight difference in age means and female sex proportions, as expected (since we simulated the missingness mechanism ourselves). After inverse intensity of visit weighting, most differences vanished, with age means, depressive mood means, and female proportions that are much closer (Suppl. Table 3).
\begin{table}[H]
\caption{ Estimates (95\% bootstrap percentiles confidence intervals) of the marginal effect of counselling on the average number of alcoholic beverages consumed, \textit{Add Health} study, United States, 1996-2008}
\centering 
\begin{tabular}{|c|c|c|c| } 
\hline 
OLS & IPT$^\phi$ & IIV$^\dagger$ & IIV$^\ddagger$  \tabularnewline
\hline
 0.62 (0.39, 0.75) & 0.34 (0.15, 0.48) & 0.64 (0.40, 0.77) & 0.72 (0.49, 0.92)   \tabularnewline
\hline
\hline
 FIPTM$^{\phi,\dagger}$ &  FIPTM$^{\phi,\ddagger}$&   AAIIW$^{\phi,\dagger}$ &  AAIIW$^{\phi,\ddagger}$ \tabularnewline
 \hline
  0.35 (0.12, 0.50) & 0.46 (0.24, 0.67)&0.35 (0.12, 0.51) & 0.28 (0.04, 0.54) \tabularnewline
 \hline 
\end{tabular}

\small{
Acronyms: IPT, inverse probability of treatment; IIV: inverse intensity of visit; FIPTM: the flexible inverse probability of treatment and monitoring; AAIIW: the doubly augmented, doubly inverse weighted.\\
$^\phi$. Note we do not know the true data-generating mechanism for the treatment mechanism in the application.\\
$^\dagger$. This estimator uses a correctly specified generating mechanism for outcome missingness. $^\ddagger$.  This estimator uses a wrongly specified generating mechanism for outcome missingness.}
\end{table}
 In this application using data from the \textit{Add Health} study, both the adjustment for confounding and the one for outcome missingness bring the estimates for the marginal effect of exposure to counselling towards the null. For instance, the inverse probability of treatment-weighted estimator is closer to $0$ than the standard ordinary least squares estimator that does not adjust for confounding or the observation process. The estimator using the correctly adjusted inverse intensity of visit weights (IIV$^{\dagger}$) slightly brings the estimate towards the null when compared with using the wrong inverse intensity of visit weights (IIV$^{\ddagger}$). Combining both adjustments, the estimate for the causal marginal effect of counselling on alcohol consumption goes from $0.62$ ($95\%$ confidence interval $0.39$, $0.75$) with no adjustment at all, to $0.35$ ($0.12$, $0.51$) when using the correct inverse intensity of visit weights (and a propensity score based on our assumptions) in the multiply robust AAIIW estimator.  

Of most interest is the comparison of the FIPTM and the AAIIW estimators in this application. Using a wrong model for the calculation of inverse intensity of visit weights leads to an estimate of $0.46$ ($95\%$ confidence interval $0.24$, $0.67$) with the FIPTM and $0.28$ ($95\%$ confidence interval $0.04$, $0.54$) with the AAIIW estimator. Both estimators lead to similar estimates when using the correctly specified inverse intensity of visit weights. Those results indicate that in a setting in which we would not know the true observation mechanism, the AAIIW estimator might still lead to an estimate of the causal effect closer to the null (around $0.2$ or $0.3$ here) while we know that the FIPTM risks being biased when its weights are not correctly modelled. This indicates that the true effect is probably closer to $0.2$ or $0.3$ in this application (and indeed, the $95\%$ confidence interval for the AAIIW estimator using the wrong inverse intensity of visit weights, which may be protected against bias since its outcome mean model conditional on the observation predictors has a chance at being correctly specified, is realistic with a lower bound closer to $0$).


\section{Discussion}\label{sec5}

This work proposed the first quadruply robust estimator that is consistent when only two out of four nuisance models (the treatment, the visit, and two conditional outcome mean models) are correctly specified, as long as the combination of correctly specified models is one shown in Table \ref{tab1}. The proposed estimator is particularly useful for observational settings subject to confounding and irregular observation times when researchers suspect they can specify correctly at least one of two models related to confounders (either the prescription mechanism, or the outcome biological mechanism), and one of two models related to the outcome observation (either the observation of the outcome or the outcome biological mechanism). The estimator also allows mediators and other variables that could be on the causal path from the treatment to the outcome to affect the observation model and yet be adjusted for. 

In addition to being more robust than its only alternative, the FIPTM which is a doubly-weighted estimator, the AAIIW estimator is also the most efficient estimator in its class of semiparametric estimators. In simulation studies, the AAIIW was demonstrated to be robust and empirically as efficient as the FIPTM when the two weight models are correctly specified but much more efficient in other scenarios (such as when all four models used in its estimating equations are correctly specified or when the two outcome mean models conditional on confounders or observation predictors are correctly specified).


In an application to the \textit{Add Health} study in the United States, we have found a difference between more naive estimators and the multiply robust AAIIW estimator in the estimation of the causal marginal effect of therapy counselling on alcohol consumption (e.g., a causal effect of 0.35 more drinks with counselling therapy (95\% confidence interval 0.12, 0.51) versus 0.62 more drinks (95\% confidence interval 0.39, 0.75) with the most naive estimator). It is possible that unmeasured confounding remains and that a better adjustment would have brought these estimates even closer to the null. Sensitivity analyses could be used to assess the effect of unmeasured confounding or visit predictors that were not accounted properly in the estimator (see \cite{smith2022trials} for informative observation or see e.g., \cite{schneeweiss2006sensitivity, vanderweele2011bias} for unmeasured confounding).

The consistency of our proposed estimator relies on specific combinations of correctly specified nuisance models listed in Table \ref{tab1} and some classical causal assumptions mentioned in $\mathsection$\ref{sec2}. An analyst using the novel approach, in collaboration with an expert from the substantive research field, should identify the confounders of the relationship between the exposure and outcome and the outcome observation predictors at risk of creating spurious associations between the exposure and the outcome (conditional exchangeability assumption). The use of a causal diagram can help in depicting the open paths by which dependencies that are not due to causal effects arise. It is not enough to include these variables in the weight models (inverse probability of treatment or inverse intensity of visit weights) or in the conditional outcome mean models discussed in this manuscript. A model that is correctly specified implies that the functional format of the predictors in the model is correctly specified. Second, the analyst should ensure that the observations included in the analysis meet the two positivity assumptions for the treatment and the observation models. They might decide to remove from the analysis a patient with a set of characteristics that are only represented in one of the two exposure groups (otherwise, the positivity of treatment assumption may be violated) or with a set of characteristics that are only represented when the outcome is not observed (otherwise violating the positivity of observation assumption). Third, the trivial causal assumption of consistency of the outcome must be met. In our application to the \textit{Add Health} study, for instance, we assume that the outcome observed in those who claimed to have received counselling is truly equal to their potential outcome under counselling, and vice-versa. Measurement errors, or patients not having filled in the information properly or truthfully could alter the estimates.

Interesting future avenues of work include the use of more flexible methods, such as machine learning methods, to correctly model the outcome mean as a function of confounders or observation predictors. The framework of targeted maximum likelihood estimation \citep{van2006targeted, schuler2017targeted} could be used for that purpose. Another advantage of the proposed estimator is that it is based on the general framework of generalized estimating equations \citep{zeger1986longitudinal} and could therefore be extended straightforwardly to account for other types of outcomes (e.g., binary outcomes). The extension to categorical or continuous exposure is also possible. 

\section*{Acknowledgement}
We thank Professor Marie Davidian at North Carolina State University for the enriching discussions we had as part of a virtual research internship of author JC in the Department of Statistics at North Carolina State University in 2021-2022.  

This research uses data from Add Health, funded by grant P01 HD31921 (Harris) from the Eunice Kennedy Shriver National Institute of Child Health and Human Development (NICHD), with cooperative funding from 23 other federal agencies and foundations. Add Health is currently directed by Robert A. Hummer and funded by the National Institute on Aging cooperative agreements U01 AG071448 (Hummer) and U01AG071450 (Aiello and Hummer) at the University of North Carolina at Chapel Hill. Add Health was designed by J. Richard Udry, Peter S. Bearman, and Kathleen Mullan Harris at the University of North Carolina at Chapel Hill.

\section*{Supplementary material}

Supplementary material available at the end of this document includes:
\begin{itemize}
\item[--] Supplementary material A: Proof of the multiple robustness of the novel estimator under the different scenarios presented in Table \ref{tab1}.
\item[--]  Supplementary material B: Sketch-proof for the efficiency of the AAIIW estimator.
\item[--]  Supplementary material C: Details of the simulation studies.
\item[--] Supplementary material D: Results of the simulation studies for a sample size of 250 with the nonhomogeneous Poisson rate for the observation indicators, for sample sizes 250 or 1000 using the Bernoulli probability for the observation of the outcome, and Monte Carlo bias and mean square error of each estimator, in all scenarios tested.
\item[--] Supplementary material E: Tables of characteristics in the \textit{Add Health} study stratified by two weighting strategies.
\end{itemize} 


\bibliographystyle{biometrika}
\bibliography{biblio_thesis}
 
  \appendix

\captionsetup[table]{name=Supplementary Table}
\captionsetup[figure]{name=Supplementary Figure}
 
  \newpage
  
\noindent \textbf{Supplementary Material for ``Quadruply robust estimation of marginal structural models in observational studies subject to covariate-driven observations''} 
 
\newpage
 
\noindent \textbf{Supplementary Material A} \vspace{0.2cm}\\

\noindent \textbf{Proof of the multiple robustness of the novel estimator under the different scenarios presented in Table 2 in the main manuscript} \vspace{0.3cm}\\
 
 \noindent The probability limiting estimating equations of the AAIIW estimator are

 \begin{eqnarray*}
 &  & \pr \left[\int_{0}^{\tau}\frac{\frac{\bone\{ A_{i}(t)=a\}}{\pr\{A_{i}(t)=a\mid K_{i}(t);\hat{\psi}\}}Y_{i}(t){\color{black}-\frac{\bone\{A_{i}(t)=a\}-\pr\{A_{i}(t)=a\mid  K_{i}(t);\hat{\psi}\}}{\pr\{A_{i}(t)=a\mid  K_{i}(t);\hat{\psi}\}}\mu_{a}\{ K_{i}(t);\hat{\alpha}_{K}\}}-\zeta_{i}(t;\beta_{a})}{\E\left\{\de N_i(t)=1\mid V_i(t);\hat{\gamma}\right\}}\de N_{i}(t)\right] \nonumber \\
 &  & - \pr \left[\color{black}\int_{0}^{\tau}\frac{\de M_{i}(t) \left( \frac{  \mu_{a}\{V_{i}(t);\hat{\alpha}_{V}\} }{\pr\{A_{i}(t)=a\mid K_{i}(t);\hat{\psi}\}}  -\frac{\bone\{A_{i}(t)=a\}-\pr\{A_{i}(t)=a\mid K_{i}(t);\hat{\psi}\}}{\pr\{A_{i}(t)=a\mid K_{i}(t);\hat{\psi}\}}\mu_{a}\{K_{i}(t);\hat{\alpha}_{K}\} -\zeta_{i}(t;\beta_{a})\right)}{\E\left\{\de N_i(t)=1\mid V_i(t);\hat{\gamma}\right\}} \right]=0. \nonumber
\end{eqnarray*}

 \noindent In this proof, denote the correct models using an asterisk onto the parameters, i.e., the correct model for the treatment is denoted by $\pr\left\{A_i(t)=a\mid K_i(t);  \psi^{*}\right\} $.\vspace{0.1cm}\\

 \noindent Under scenario a) (Table 2) we have

 \begin{eqnarray*}
 &  & \pr \left[\int_{0}^{\tau}\frac{\frac{\bone\{ A_{i}(t)=a\}}{\pr\{A_{i}(t)=a\mid K_{i}(t);\psi^{*}\}}Y_{i}(t){\color{black}-\frac{\bone\{A_{i}(t)=a\}-\pr\{A_{i}(t)=a\mid  K_{i}(t);\psi^* \}}{\pr\{A_{i}(t)=a\mid  K_{i}(t); \psi^*\}}\mu_{a}\{ K_{i}(t);\hat{\alpha}_{K}\}}-\zeta_{i}(t;\beta_{a})}{\pr\left\{\de N_i(t)=1\mid V_i(t);\gamma^* \right\} }\de N_{i}(t)\right]   \\
 &  & - \pr \left[\color{black}\int_{0}^{\tau}\frac{ (\de N_i(t) -\pr\left\{\de N_i(t)=1\mid V_i(t); \gamma^* \right\}  ) \left( \frac{  \mu_{a}\{V_{i}(t);\hat{\alpha}_{V}\} }{\pr\{A_{i}(t)=a\mid K_{i}(t); \psi^*\}}  -\frac{\bone\{A_{i}(t)=a\}-\pr\{A_{i}(t)=a\mid K_{i}(t); \psi^*\}}{\pr\{A_{i}(t)=a\mid K_{i}(t); \psi^*\}}\mu_{a}\{K_{i}(t);\hat{\alpha}_{K}\} \right) }{\pr\left\{ \de N_i(t)=1 \mid V_i(t); \gamma^* \right\} } \right]   \\
 & & +  \pr \left[\color{black}\int_{0}^{\tau}\frac{ (\de N_i(t) -\pr\left\{\de N_i(t)=1\mid V_i(t); \gamma^*\right\} ) \zeta_{i}(t;\beta_{a})  }{\pr\left\{\de N_i(t)=1\mid V_i(t); \gamma^*\right\} } \right] \\
 & & = \pr \left[  \int_{0}^{\tau}   \frac{  \frac{\bone\{ A_{i}(t)=a\}}{\pr\{A_{i}(t)=a\mid K_{i}(t);\psi^{*}\}} Y_{i}(t) -\zeta_{i}(t;\beta_{a})}{\pr\left\{ \de N_i(t)=1\mid V_i(t);\gamma^*  \right\} } \de N_{i}(t) \right] \\
 & & = 0
\end{eqnarray*}
which is clearly unbiased for the parameter of interest since both weights are correctly specified in scenario a). \vspace{0.2cm}\\
In scenario b) (Table 2), we rewrite the original estimating equations as follows:

 \begin{eqnarray*}
 & &  \pr \left[\int_{0}^{\tau}\frac{\frac{\bone\{ A_{i}(t)=a\}}{\pr\{A_{i}(t)=a\mid K_{i}(t);\hat{\psi}\}}(  Y_{i}(t){\color{black}- \mu_{a}\{ K_{i}(t);\hat{\alpha}_{K}\}} ) + \mu_{a}\{ K_{i}(t);\hat{\alpha}_{K}\} -\zeta_{i}(t;\beta_{a})}{\E\left\{\de N_i(t)=1\mid V_i(t);\hat{\gamma}\right\}}\de N_{i}(t)\right] \nonumber \\
& & - \pr \left[\int_{0}^{\tau}\frac{\frac{\bone\{ A_{i}(t)=a\}}{\pr\{A_{i}(t)=a\mid K_{i}(t);\hat{\psi}\}}(  \mu_{a}\{ V_{i}(t);\hat{\alpha}_{V}\}{\color{black}- \mu_{a}\{ K_{i}(t);\hat{\alpha}_{K}\}} ) + \mu_{a}\{ K_{i}(t);\hat{\alpha}_{K}\} -\zeta_{i}(t;\beta_{a})}{\E\left\{\de N_i(t)=1\mid V_i(t);\hat{\gamma}\right\}}\de N_{i}(t)\right] \nonumber \\
& & +\pr \left[\int_{0}^{\tau}\frac{\frac{\bone\{ A_{i}(t)=a\}}{\pr\{A_{i}(t)=a\mid K_{i}(t);\hat{\psi}\}}(  \mu_{a}\{ V_{i}(t);\hat{\alpha}_{V}\}{\color{black}- \mu_{a}\{ K_{i}(t);\hat{\alpha}_{K}\}} ) + \mu_{a}\{ K_{i}(t);\hat{\alpha}_{K}\} -\zeta_{i}(t;\beta_{a})}{\E\left\{\de N_i(t)=1\mid V_i(t);\hat{\gamma}\right\}} \E\left\{\de N_i(t)=1\mid V_i(t);\hat{\gamma}\right\}\right]. \nonumber \\
\end{eqnarray*}
Then, using the asterisk notation for the correct models once again, we obtain

 \begin{eqnarray*}
 & & \pr \left[\int_{0}^{\tau}\frac{\frac{\bone\{ A_{i}(t)=a\}}{\pr\{A_{i}(t)=a\mid K_{i}(t);\hat{\psi}\}}(  Y_{i}(t){\color{black}- \mu_{a}\{ K_{i}(t); \alpha^*_K\}} ) + \mu_{a}\{ K_{i}(t); \alpha^*_K \} -\zeta_{i}(t;\beta_{a})}{\E\left\{\de N_i(t)=1\mid V_i(t);\hat{\gamma}\right\}}\de N_{i}(t)\right]  \\
& & - \pr \left[\int_{0}^{\tau}\frac{\frac{\bone\{ A_{i}(t)=a\}}{\pr\{A_{i}(t)=a\mid K_{i}(t);\hat{\psi}\}}(  \mu_{a}\{ V_{i}(t); \alpha^*_V\}{\color{black}- \mu_{a}\{ K_{i}(t); \alpha^*_K\}} ) + \mu_{a}\{ K_{i}(t); \alpha^*_{K}\} -\zeta_{i}(t;\beta_{a})}{\E\left\{\de N_i(t)=1\mid V_i(t);\hat{\gamma}\right\}}\de N_{i}(t)\right]  \\
& & +\small{\pr \left[\int_{0}^{\tau}\frac{\frac{\bone\{ A_{i}(t)=a\}}{\pr\{A_{i}(t)=a\mid K_{i}(t);\hat{\psi}\}}(  \mu_{a}\{ V_{i}(t); \alpha^*_{V}\}{\color{black}- \mu_{a}\{ K_{i}(t); \alpha^*_{K}\}} ) + \mu_{a}\{ K_{i}(t); \alpha^*_{K}\} -\zeta_{i}(t;\beta_{a})}{\E\left\{\de N_i(t)=1\mid V_i(t);\hat{\gamma}\right\}} \E\left\{\de N_i(t)=1\mid V_i(t);\hat{\gamma}\right\}\right] }  \\
& & = \pr \left[ \int_{0}^{\tau}  \frac{   \frac{  \bone\{ A_{i}(t)=a\}}{\pr\{A_{i}(t)=a\mid K_{i}(t);\hat{\psi}\} } (  Y_{i}(t){ - \mu_{a}\{ V_{i}(t); \alpha^*_V\}} )  } { \E \left\{ \de N_i(t)=1\mid V_i(t);\hat{\gamma}\right\} } \de N_{i}(t)\right]   \text{(by combining the first two lines).} \\
\end{eqnarray*}
The last equation is unbiased since the model $\mu_{a}\{ V_{i}(t); \alpha^*_V\}$ is correctly specified . \vspace{0.2cm}\\

\noindent In scenario c) (Table 2), we have the following equations once plugging-in the correct specified models:

 \begin{eqnarray*}
 &  & \pr \left[\int_{0}^{\tau}\frac{\frac{\bone\{ A_{i}(t)=a\}}{\pr\{A_{i}(t)=a\mid K_{i}(t);\hat{\psi}\}} \left[ Y_{i}(t) -\mu_{a}\{ K_{i}(t); \alpha^*_{K}\} \right] + \mu_{a}\{ K_{i}(t); \alpha^*_{K}\}
-\zeta_{i}(t;\beta_{a})}{\pr\left\{\de N_i(t)=1\mid V_i(t); \gamma^* \right\}}\de N_{i}(t)\right] \nonumber \\
 &  & -   \pr  \color{black}\int_{0}^{\tau}\frac{ \de N_i(t)-\pr\left\{\de N_i(t)=1\mid V_i(t); \gamma^*\right\} }{\pr\left\{\de N_i(t)=1\mid V_i(t); \gamma^*\right\}} 
  \left( \frac{  \mu_{a}\{V_{i}(t);\hat{\alpha}_{V}\} }{\pr\{A_{i}(t)=a\mid K_{i}(t);\hat{\psi}\}} \right) \\
  & & \small{ + \pr  \color{black}\int_{0}^{\tau}\frac{ \de N_i(t)-\pr\left\{\de N_i(t)=1\mid V_i(t); \gamma^*\right\} }{\pr\left\{\de N_i(t)=1\mid V_i(t); \gamma^*\right\}}  \left( \frac{\bone\{A_{i}(t)=a\}-\pr\{A_{i}(t)=a\mid K_{i}(t);\hat{\psi}\}}{\pr\{A_{i}(t)=a\mid K_{i}(t);\hat{\psi}\}}\mu_{a}\{K_{i}(t); \alpha^*_{K}\} -\zeta_{i}(t;\beta_{a})\right) }  \\
  & & =  \pr \left[\int_{0}^{\tau} \frac{ \mu_{a}\{ K_{i}(t); \alpha^*_{K}\}
-\zeta_{i}(t;\beta_{a})}{\pr\left\{\de N_i(t)=1\mid V_i(t); \gamma^* \right\}}\de N_{i}(t)\right] =0. \nonumber \nonumber
\end{eqnarray*}

\noindent The two last rows in the first development above cancel out because the martingale residuals are 0-mean under correctly specified IIV weights. The last equation above is unbiased for the causal effect since the outcome model conditional on the confounders and the IIV weights are correctly specified. \vspace{0.2cm}\\

\noindent Finally, in scenario d), we transform the original equations into:

 \begin{eqnarray*}
 &  & \pr \left[\int_{0}^{\tau}\frac{\frac{\bone\{ A_{i}(t)=a\}}{\pr\{A_{i}(t)=a\mid K_{i}(t);\hat{\psi}\}} \left[ Y_{i}(t)- \mu_{a}\{V_{i}(t);\hat{\alpha}_{V}\}\right]   }{\E\left\{\de N_i(t)=1\mid V_i(t);\hat{\gamma}\right\}}\de N_{i}(t)\right] \nonumber \\
 &  & + \pr \left[\color{black}\int_{0}^{\tau}  \left( \frac{ \bone\{A_{i}(t)=a\}    }{\pr\{A_{i}(t)=a\mid K_{i}(t);\hat{\psi}\}}  \left[\mu_{a}\{V_{i}(t);\hat{\alpha}_{V}\} -\mu_{a}\{K_{i}(t);\hat{\alpha}_{K}\} \right]  + \mu_{a}\{K_{i}(t);\hat{\alpha}_{K}\} -\zeta_{i}(t;\beta_{a}) \right) \de  t \right]=0. \nonumber
\end{eqnarray*}

Once replacing with the correctly specified models in scenario d), we obtain

\begin{eqnarray*}
 &  & \pr \left[\int_{0}^{\tau}\frac{\frac{\bone\{ A_{i}(t)=a\}}{\pr\{A_{i}(t)=a\mid K_{i}(t); \psi^*\}} \left[ Y_{i}(t)- \mu_{a}\{V_{i}(t); \alpha^*_{V}\}\right]   }{\E\left\{\de N_i(t)=1\mid V_i(t);\hat{\gamma}\right\}}\de N_{i}(t)\right] \nonumber \\
 &  & + \pr \left[\color{black}\int_{0}^{\tau}  \left( \frac{ \bone\{A_{i}(t)=a\}    }{\pr\{A_{i}(t)=a\mid K_{i}(t); \psi^*\}}  \left[\mu_{a}\{V_{i}(t); \alpha^*_{V}\} -\mu_{a}\{K_{i}(t);\hat{\alpha}_{K}\} \right]  + \mu_{a}\{K_{i}(t);\hat{\alpha}_{K}\} -\zeta_{i}(t;\beta_{a}) \right) \de  t \right] \\
 & & = \pr \left[\int_{0}^{\tau}\frac{\frac{\bone\{ A_{i}(t)=a\}}{\pr\{A_{i}(t)=a\mid K_{i}(t); \psi^*\}} \left[ Y_{i}(t)- \mu_{a}\{V_{i}(t); \alpha^*_{V}\}\right]   }{\E\left\{\de N_i(t)=1\mid V_i(t);\hat{\gamma}\right\}}\de N_{i}(t)\right] \nonumber \\
 & & + \pr \left[\color{black}\int_{0}^{\tau}  \left( \frac{ \bone\{A_{i}(t)=a\}    }{\pr\{A_{i}(t)=a\mid K_{i}(t); \psi^*\}}   \mu_{a}\{V_{i}(t); \alpha^*_{V}\} \right) dt \right] \\
 & & - \pr \left[ \left( \frac{ \bone\{A_{i}(t)=a\}  -\pr\{A_{i}(t)=a\mid K_{i}(t); \psi^*\}   }{\pr\{A_{i}(t)=a\mid K_{i}(t); \psi^*\}}  \mu_{a}\{K_{i}(t);\hat{\alpha}_{K}\}     +\zeta_{i}(t;\beta_{a}) \right) \de  t \right] \\
 & & = \pr \left[\color{black}\int_{0}^{\tau}  \left( \frac{ \bone\{A_{i}(t)=a\}    }{\pr\{A_{i}(t)=a\mid K_{i}(t); \psi^*\}}   \mu_{a}\{V_{i}(t); \alpha^*_{V}\} - \zeta_{i}(t;\beta_{a}) \right) dt \right] =0. \nonumber
\end{eqnarray*}
The last equation above is unbiased for the causal effect.

 \newpage

\noindent \textbf{Supplementary Material B} \vspace{0.2cm}\\
\noindent \textbf{Variance estimation and proof of efficiency of the AAIIW estimator compared to the FIPTM.}\vspace{0.2cm}\\
   
\noindent To make the demonstrations lighter, denote  \begin{align*}
     I_a = & \bone\{A_i(t)=a\} \\ 
     e_a =& pr\{A_{i}(t)= a \mid K_{i}(t); \psi_0\}\\
     \de N =& \bone\{\de N_i(t)=1\} \\
     \rho =& \E \{\de N_{i}(t)=1\mid V_{i}(t); \gamma_0\} \\
     \de M =& \de N_i(t)- \rho  \\
     \mu_{aK} =& \mu_{a} \{  K_i(t); \alpha_{0K} \}\\
     \mu_{aV} = &\mu_{a} \{  V_i(t); \alpha_{0V} \} \\
     \mu_0 =& \E[Y^0_i(t)] = \beta_0 \\
     \mu_1 =& \E[Y^1_i(t)] = \beta_0 + \beta_1.
\end{align*}

\noindent With the simpler notation, and under all nuisance models correctly specified, the FIPTM (in this case, we do not consider modelling flexibly the intercept as in the original paper proposing the FIPTM but rather consider a standard doubly weighted least squares estimator for the FIPTM) and AAIIW estimators respectively correspond to the solutions of the following sets of estimating equations:
\begin{align}
\E_{n}\left[\int_{0}^{\tau} \frac{\de N}{\rho} \frac{ I_a }{e_a} \left\{ Y_{i}(t)-\zeta_{i}(t;\beta_{a})\right\}\right] =0 \label{eqt}
\end{align}
and
\begin{eqnarray}
 &  & \E_n \left[\int_{0}^{\tau} \left\{ \frac{ I_a}{e_a}Y_{i}(t){ - \left(\frac{I_a- e_a }{e_a}\right) \mu_{aK}}-\zeta_{i}(t;\beta_{a})   \right\} \frac{\de N}{\rho}  \right]   -  \E_n \left[\color{black}\int_{0}^{\tau} \frac{\de M}{\rho}  \left\{ \frac{ \mu_{aV} }{ e_a}  - \left(\frac{I_a- e_a }{e_a}\right)\mu_{aK} - \zeta_{i}(t;\beta_{a})\right\}   \right] \nonumber \\
 & & =0.  
\end{eqnarray}
Note that each set of estimating equations contains two equations, one for $\beta_0$ and one for $\beta_1$ once $\beta_0$ is evaluated. We use the influence function of each estimator to derive its asymptotic variance. The variance of each estimator equals to the variance of its influence function. \\

\noindent Denote by $\beta_a= [\beta_0 \hspace{0.1cm} \beta_1]$ the vector of true parameters. We prove the relative efficiency of the AAIIW estimator when compared with the FIPTM estimator under the assumption that all nuisance models are correctly specified for both estimators. The influence function of the $\hat{\beta}_{DW}$ estimator, denoted $\psi(o_i)= \psi(o_i; \beta_a, \delta_0)$ for $o_i$ the observation data of the ith individual, and $\delta_0=\left\{ \psi_0, \gamma_0 \right\}$ the true nuisance parameters (for the treatment and observation models) must satisfy
$$\sqrt{n}\left(\hat{\beta}_{DW} - \beta_a \right) = \frac{1}{\sqrt{n}} \sum_{i=1}^n \psi(o_i) + \smallO(1).$$

\noindent By replacing terms in equation \ref{eqt} above, we obtain a vector of influence functions the same size as the vector of parameters of interest $\beta_a$, that is:

$$\psi(o_i) = \left[ \begin{matrix} \int_{t=0}^{\tau} \frac{ \de N}{ \rho}   \frac{ I_0}{ e_0}  \left\{ Y_{i}(t) - \mu_0  \right\} \\  \int_{t=0}^{\tau} \frac{\de N }{ \rho}   \left\{   \left(   \frac{ I_1}{e_1}  -  \frac{ I_0}{ e_0} \right) Y_{i}(t) + \mu_0 I_0 - \mu_1 I_1     \right\} \end{matrix} \right].  $$

\noindent For the variance of the DW estimator, our interest is in $\beta_1$ and so we focus on the second equation (the only equation that depends on it), assuming we already have an estimate for $\beta_0$ from the first equation. Using similar developments as those that follow, we could show that the variance of the estimate for $\beta_0$ is also smaller when using the AAIIW estimator as compared with the FIPTM estimator, but this is omitted in what follows.\\

\noindent The variance of the FIPTM estimator equals to the variance of  its influence function. We denote that variance by $\sigma_{FIPTM}^2$. In further derivations, we drop the integral that sums the terms over all times when there is an observation of the outcome (as this integral sum is the same for the FIPTM and the AAIIW estimator). We, therefore, focus on only one term in the integral, yielding:

\begin{align}
\sigma_{FIPTM}^2 =& \E [ \psi(o_i)^2] + \E[\psi(o_i)]^2 \nonumber \\
 =&  \E [ \psi(o_i)^2] + 0\nonumber \\
=& \E \left(  \left[ \frac{ \de N}{ \rho} \left\{   \left(   \frac{ I_1 }{ e_1}  -  \frac{I_0}{ e_0} \right) Y_{i}(t) + \mu_0 I_0  - \mu_1 I_1   \right\} \right] ^2 \right) \nonumber \\
  =& \E \left[  \overbrace{  \left\{ \frac{\de N}{\rho }       \left(   \frac{I_1}{e_1}   \right)  ( Y_{i}(t)  - \mu_1   )\right\}^2   }^{A} \right]  + \E \left[  \overbrace{ \left\{ \frac{\de N}{\rho }        \left(   \frac{I_0}{e_0}   \right) ( Y_{i}(t) - \mu_0  )\right\}^2 }^{B} \right]    - 2\E  \left[ \E \left\{ A \times B \mid V_i(t) \right\}\right]  \nonumber\\
=& \E \left[     \frac{\de N}{\rho^2}          \frac{I_1}{e_1^2}     ( Y_{i}(t) -  \mu_1  )^2 \right]  + \E \left[    \frac{\de N }{\rho^2 }         \frac{ I_0}{ e_0^2}     ( Y_{i}(t) - \mu_0    )^2 \right]  - 2\E  \left\{ \E \left[ A \mid V_i(t) \right] \times  \E \left[ B \mid V_i(t) \right] \right\} \nonumber \\
&\text{with the third term true since both terms are independent given V (treated and untreated patients). Now, }\nonumber \\
&\text{the third term is 0 because each expectation conditional on $V_i(t)$ is zero.} \nonumber\\
& =\E \left[     \frac{ \E [\de N \mid V_i(t) ] }{\rho^2}          \frac{\E [I_1 \mid V_i(t) ] }{e_1^2}    \E\left[ ( Y_{i}(t)   - \mu_1  )^2  \mid V_i(t), A_i(t)=1, \de N_i(t)=1 \right] \right]  \nonumber \\
&\hspace{0.4cm}  + \E \left[    \frac{\E [\de N \mid V_i(t) ]  }{\rho^2 }         \frac{ \E [I_0 \mid V_i(t) ]  }{ e_0^2}      \E\left[ ( Y_{i}(t) - \mu_0    )^2  \mid V_i(t) , A_i(t)=0, \de N_i(t)=1\right] \right]  + 0 \nonumber \\
& =\E \left[     \frac{ 1}{\rho }          \frac{1 }{e_1 }     \E\left[ ( Y_{i}(t)   - \mu_1  )^2  \mid V_i(t), A_i(t)=1, \de N_i(t)=1 \right] \right] \nonumber\\
&\hspace{0.4cm}+ \E \left[    \frac{1 }{\rho  }         \frac{ 1}{ e_0 }      \E\left[ ( Y_{i}(t) - \mu_0    )^2  \mid V_i(t) , A_i(t)=0, \de N_i(t)=1\right] \right]   \nonumber \\
& =\E \left[  \frac{   (Y^1_i(t) - \mu_1)^2  }{\rho e_1 } \right]         + \E \left[  \frac{  (Y^0_i(t) - \mu_0)^2  }{\rho e_0 } \right].    \label{ab}
\end{align}

 \noindent For the AAIIW estimator, we need to show that the variance is smaller than (\ref{ab}). First we derive the influence function $\phi(o_i)=\phi(o_i; \beta_a, \delta_0)$ for $o_i$ the observation data of the ith individual and $\delta_0$ now defined as the vector of true parameters $\left\{ \psi_0, \gamma_0, \alpha_{K0}, \alpha_{V0} \right\}$ (which now includes the parameters of the two conditional outcome mean models). We have that the influence function satisfies
$$\sqrt{n}(\hat{\beta}_{AAIIW}-\beta_a) = \frac{1}{\sqrt{n}} \sum_{i=1}^n \phi(o_i) + \smallO(1)$$
or, using the formula for M-estimators provided in Tsiatis (Chapter 3, section 3.2) and noting that the AAIIW estimator is an M-estimator and is the solution to 

\begin{align}
    \sum_{i=1}^n m(o_i; \hat{\beta}_{AAIIW} ) = 0 \label{ac}
\end{align}
with the m-function that can be derived directly from (\ref{ac}), then we can compute the influence function by first computing the partial derivatives of the $m$ function w. r. t. to the two parameters in $\beta_a$ (use, e.g., formula 3.6 in Tsiatis) and find that

\begin{align*}
\phi(o_i) =& \psi(o_i) + \left[ \begin{matrix}  \int_{t=0}^{\tau} -\frac{\de N }{\rho}  \left(\frac{ I_0- e_0  }{ e_0} \right) \mu_{0K} -\frac{\de M }{\rho}  \left\{  \frac{ \mu_{0V}  }{ e_0} -  \left(\frac{I_0- e_0  }{ e_0} \right) \mu_{0K}   - \mu_0   \right\}   \\  \int_{t=0}^{\tau} -\frac{\de N }{ \rho}\left(\frac{ I_1 - e_1  }{ e_1} \right) \mu_{1K}   -\frac{\de M }{\rho}\left\{\frac{\mu_{1V}   }{e_1}  -  \left(\frac{I_1- e_1 }{ e_1} \right) \mu_{1K} -  \mu_1   \right\}   \\
  \hspace{1cm}  + \frac{\de N }{ \rho}\left(\frac{ I_0 - e_0  }{ e_0} \right) \mu_{0K}  +\frac{\de M }{\rho}\left\{\frac{\mu_{0V}   }{e_0}  -  \left(\frac{I_0- e_0 }{ e_0} \right) \mu_{0K} -\mu_0   \right\}  \end{matrix} \right]    
\end{align*}
with the second part added to $\psi(o_i)$ corresponding to the augmented terms. Note that the second row of $\phi(o_i)$ contains two terms, one being the augmented term for the treated and one for the untreated. The change in sign for both terms is due to the subtraction between the augmented terms due to the treated and the untreated patients. \\

\noindent To compute the variance of the AAIIW, that we denote by $\sigma_{AAIIW}^2$, we compute the variance of its influence function. We again focus on the second influence function assuming that $\beta_0$ is already estimated, and again we drop the integral sign in the following derivations. We have:

\begin{align*}
\sigma_{AAIIW}^2 =& \E [ \phi(o_i)^2]   \\
  =& \E\left[ \psi(o_i)^2 \right] \\
 &\hspace{0.4cm}    + \E\left[   \left(\overbrace{ - \left\{\frac{ I_1 - e_1  }{ e_1 } \right\} \mu_{1K} - \frac{\de M}{\rho} \left\{ \frac{\mu_{1V}}{e_1}  -    \mu_1   \right\}  }^{C_1}   +  \overbrace{     \left\{\frac{ I_0 - e_0  }{ e_0 } \right\} \mu_{0K} +\frac{\de M}{\rho} \left\{ \frac{\mu_{0V}}{e_0} -      \mu_0    \right\}   }^{C_2}  \right)^2 \right]  \\
 &\hspace{1.2cm} + 2 \E[\psi(o_i) \times (C_1 + C_2)].
 \end{align*}

\noindent The first term, $\E\left[ \psi(o_i)^2 \right] $, is the variance of the FIPTM estimator. Denote the second and third terms by $D= \E[ (C_1+C_2)^2]$ and $F=  2 \E[\psi(o_i) \times (C_1+C_2)]$. We must show that $D+F < 0$ to show relative efficiency of the AAIIW estimator.  We have that

\begin{align*}
D = \E[(C_1+C_2)^2] = & \E[C_1^2 +C_2^2 + 2C_1C_2]  = \E[C_1^2] +\E[C_2^2] + \E[2C_1C_2] 
 \end{align*}
 where
 \begin{align*}
 \E[C_1^2]   = & \E \left[     \frac{\de M^2}{\rho^2}  \left( \frac{\mu_{1V}}{e_1} -  \mu_1   \right)^2 \right] + \E \left[    \left(\frac{I_1 - e_1  }{ e_1 } \right)^2 \mu^2_{1K}   \right]\\
 & -2 \E \left[ \E \left[ - \frac{\de M}{\rho} \left\{ \frac{\mu_{1V}}{e_1} -    \mu_1  \right\}  \left( \frac{ I_1 - e_1  }{ e_1 } \right) \mu_{1K} \mid V_i(t)  \right] \right] \\
 = & \E \left[     \frac{\de M^2}{\rho^2}  \left( \frac{\mu_{1V}}{e_1} -  \mu_1  \right)^2 \right] + \E \left[    \frac{ I_1 + e_1^2 - 2 I_1 e_1  }{ e_1^2 }    \mu^2_{1K}      \right]\\
 &\text{where the third term was removed since}\hspace{0.2cm} \E \left[  \de M    \mid V_i(t) \right]=0 \\
=& \E \left[     \frac{\E[ (\de N + \rho^2 - 2 \de N \rho )\mid V_i(t) ] }{\rho^2}    \left(  \E[Y^1_i(t)] -  \mu_1   \right)^2 \  \right] + \mu_1^2  \E \left[ \frac{ \E \left[ I_1 + e_1^2 - 2 I_1 e_1 \mid V_i(t) \right] }{e_1^2}       \right]    \\
=& 0+ \mu_1^2 \E \left[   \frac{1-e_1}{e_1} \right]  \hspace{0.2cm} \text{since} \hspace{0.2cm}  \E[Y^1_i(t)] -  \mu_1  =0.
\end{align*}
Similarly, we have

\begin{align*}
 \E[C_2^2]   = & \mu_0^2 \E \left[   \frac{1-e_0}{e_0} \right]. 
 \end{align*}
 The third term, $\E[2C_1C_2]$, cancels out, after using iterated expectation and conditioning on e.g., $V_i(t)$, as the martingale residuals and the treatment residuals are zero-mean conditional on that set. Thus,

\begin{align*}
D = &  \mu_1^2 \E \left[   \frac{1-e_1}{e_1}  \right]+ \mu_0^2 \E \left[   \frac{1-e_0}{e_0}  \right] .
\end{align*}
We are left with calculting $F$. We have:
\begin{align*}
F = &  2 \E[\psi(o_i) \times (C_1+C_2)] \\
  = &  2 \E[\psi(o_i) \times  C_1 ] + 2 \E[\psi(o_i) \times  C_2 ].
\end{align*}
We start with the left term and have
\begin{align*}
 \E[\psi(o_i) \times  C_1 ] = & \E \left[ \frac{\de N }{ \rho}   \left\{   \left(   \frac{ I_1}{e_1}  -  \frac{ I_0}{ e_0} \right) Y_{i}(t) + \mu_0 I_0 - \mu_1 I_1     \right\} \times \left( - \left\{\frac{ I_1 - e_1  }{ e_1 } \right\} \mu_{1K} - \frac{\de M}{\rho} \left\{ \frac{\mu_{1V}}{e_1}  -    \mu_1   \right\} \right)\right] \\
 = & - \E \left[ \frac{\de N }{ \rho}   \left\{   \left(   \frac{ I_1}{e_1}  -  \frac{ I_0}{ e_0} \right) Y_{i}(t) + \mu_0 I_0 - \mu_1 I_1     \right\} \times \left(   \left\{\frac{ I_1 - e_1  }{ e_1 } \right\} \mu_{1K}+ \frac{\de M}{\rho} \left\{ \frac{\mu_{1V}}{e_1}  -    \mu_1   \right\} \right)\right] \\
 = &- \E \left[ \frac{\de N }{ \rho}   \left\{     \frac{ I_1}{e_1}    Y_{i}(t) - \mu_1         \right\} \times   \left\{\frac{ I_1 - e_1  }{ e_1 } \right\} \mu_{1K} \right] + \E \left[ \frac{\de N }{ \rho}   \left\{     \frac{ I_0}{e_0}    Y_{i}(t)- \mu_0         \right\} \times   \left\{\frac{ I_1 - e_1  }{ e_1 } \right\} \mu_{1K} \right] \\
 & \hspace{0.25cm} - \E \left[ \frac{\de N }{ \rho}   \left\{     \frac{ I_1}{e_1}    Y_{i}(t) - \mu_1         \right\} \times  \frac{\de M}{\rho} \left\{ \frac{\mu_{1V}}{e_1}  -    \mu_1   \right\}   \right] + \E \left[ \frac{\de N }{ \rho}   \left\{     \frac{ I_0}{e_0}    Y_{i}(t) - \mu_0         \right\} \times  \frac{\de M}{\rho} \left\{ \frac{\mu_{1V}}{e_1}  -    \mu_1   \right\}   \right] \\
  & = - \E \left[ \frac{\de N }{ \rho}   \left\{     \frac{ I_1- e_1 I_1}{e_1^2}    Y_{i}(t) - \left\{\frac{ I_1 - e_1  }{ e_1 } \right\} \mu_1         \right\} \times     \mu_{1K} \right]   \\
  & \hspace{0.5cm} + \E \left[ \frac{\de N }{ \rho}   \left\{     \frac{ I_0 (I_1 - e_1 ) }{e_0 e_1}    Y_{i}(t) - \left\{  \frac{I_1-e_1}{e_1} \right\} \mu_0  \right\}  \times \mu_{1K}\right] \\
  & \hspace{1cm} - \E \left[ \frac{\de N -\de N \rho}{ \rho^2}   \left\{     \frac{ I_1}{e_1}    Y_{i}(t) - \mu_1         \right\}  \times \left\{ \frac{\mu_{1V}}{e_1}  -    \mu_1   \right\}   \right] \\
 & \hspace{1.5cm} + \E \left[ \frac{\de N - \de N\rho }{ \rho^2}   \left\{     \frac{ I_0}{e_0}    Y_{i}(t) - \mu_0         \right\} \times    \left\{ \frac{\mu_{1V}}{e_1}  -    \mu_1   \right\}   \right].
\end{align*}
Taking iterated expectation and conditioning on $V_i(t)$, several residuals cancel out and we are left with
\begin{align*}
& = -  \E \left[  \frac{ 1- e_1}{e_1}  \mu_1^2  \right]    -\E \left[ \mu_1^2 \right] - 0 + 0 \\
& = -\mu_1^2\E \left[     \frac{1}{e_1}  \right]
 \end{align*}
 Similarly, we find
\begin{align*}
 \E[\psi(o_i) \times  C_2 ] = - \mu_0^2\E \left[     \frac{ 1 }{e_0}  
 \right].
 \end{align*}

 The variance of the AAIIW is therefore obtained by summing:
\begin{align*}
\sigma^2_{AAIIW} &= \sigma^2_{FIPTM} + D + F \\
 &= \sigma^2_{FIPTM} + \mu_1^2 \E \left[   \frac{1-e_1}{e_1}  \right]+ \mu_0^2 \E \left[   \frac{1-e_0}{e_0}  \right]
   - 2 \mu_1^2 \E \left[    \frac{ 1 }{e_1}  
 \right] - 2\mu_0^2 \E \left[   \frac{ 1 }{e_0}  
 \right]   \\
& \hspace{2cm} \iff \\
  \sigma^2_{AAIIW} - \sigma^2_{FIPTM} &= \mu_1^2 \E \left[  \frac{1-e_1}{e_1} - \frac{2}{e_1}  \right]   + \mu_0^2 \E \left[    \frac{1-e_0}{e_0}  - \frac{2}{e_0}  \right]. 
\end{align*}
And, since $\mu_1^2, \mu_0^2 > 0$, 
and $$\frac{1-e_1}{e_1}  - \frac{2}{e_1} 
= \frac{-1 -e_1}{e_1} \le \frac{-1 }{e_1} < 0 \hspace{0.2cm} \text{for all} \hspace{0.2cm} e_1 \hspace{0.2cm} \text{s. t.} \hspace{0.2cm} 0 \le e_1 \le 1$$ and 
 $$\frac{1-e_0}{e_0}  - \frac{2}{e_0} 
 = \frac{-1 -e_0}{e_0} \le \frac{-1}{e_0} <0  \hspace{0.2cm} \text{for all} \hspace{0.2cm} e_0 \hspace{0.2cm} \text{s. t.} \hspace{0.2cm} 0 \le e_0 \le 1$$
 we find that 
 $$\sigma^2_{AAIIW} - \sigma^2_{FIPTM} <0$$
$$ \iff \sigma^2_{AAIIW}<\sigma^2_{FIPTM}$$
under all nuisance models correctly specified for both estimators.
\newpage

\noindent \textbf{Supplementary Material C} \vspace{0.2cm}\\

\noindent \textbf{Simulation studies details} \\

Data were simulated to emulate a setting in which time is continuous and covariates can be measured or updated at any point in time. For that, we used a time grid starting at time 0 spanning up to time $\tau=2$ over which the variables were simulated at each 0.1-width time bin. First, the confounders, the treatment, the mediator, the pure predictor and the outcome were simulated over each of these small time bins, after which the observation process was simulated (and outcomes were removed). The individual index is omitted in what follows.

For each individual, three baseline confounders were simulated as $K_1\sim N(1,1)$,\\ $K_2\sim \text{Bernoulli}(0.55)$ and $K_3 \sim N(0,1)$ and were repeated throughout follow-up time. At each time $t$, a time-varying binary treatment was simulated as $A(t)\sim \text{Bernoulli}(p_t)$ with $p_t=\text{expit}(-0.5 + 0.8\hspace{0.1cm}K_1-0.4\hspace{0.1cm}K_2-0.4\hspace{0.1cm}K_3)$ where expit$(j)=\exp(j)/\left\{1+\exp(j)\right\}$ and the confounders are kept time-fixed (i.e., we emulate a setting in which they are measured at baseline). At each time $t$, a mediator $M(t)$ of the causal effect of $A(t)$ on the outcome $Y(t)$ was further simulated as $M(t)\mid A(t)=1 \sim N(2,1)$ and $M(t)\mid A(t)=0 \sim N(4,2)$ where the parameters in the Normal respectively correspond to the mean and variance of the random variable. A time-varying pure predictor was simulated as $P(t) \sim N(0.5, 0.09)$. 

A few details are particularly important in the simulation of the outcome $Y(t)$. First, since we would like to assess the performance of the different approaches when a mediator affects the outcome observation process, the outcome must depend on the mediator $M(t)$ and the mediator must affect the outcome observation. However, we would like to know the gold standard for the causal effect in the simulations, i.e., the true causal marginal effect of treatment once we marginalize the outcome mean model over $M(t)$. For that, we first modeled $E[M(t)\mid A(t), K_1, K_2, K_3]$ using a least squares linear model. Then, we simulated the outcome as
$$Y(t) = \kappa + 1\hspace{0.15cm} A(t) + 0.4\hspace{0.15cm} K_1 + 0.05 \hspace{0.15cm} K_2 - 0.6 \hspace{0.15cm} K_3  +$$
$$ \hspace{2cm}3 \left\{ M(t) - E[M(t)\mid A(t), K_1, K_2, K_3] \right\}+ 0.3\hspace{0.15cm}P(t) + \epsilon(t)$$
where $\kappa= 0.5$, $\epsilon(t) \sim N(\phi, 0.01)$, and $\phi \sim N(0, 0.04)$. We let $\phi$ vary by individual. When marginalizing over $M(t)$, we obtain a true marginal causal effect of 1 for treatment $A(t)$. This would not be the case if we had interactions between the treatment $A(t)$ and the confounders $K$ in the outcome generating mechanism above.

The outcome above was first simulated once for each point on the grid. The outcome values were then set to missing according to a covariate-dependent observation process with an observation rate (Poisson model) or probability (Bernoulli model) depending on the treatment, the mediator, the pure predictor, and the confounders. For each time point, this was done by simulating an observation indicator (one minus a missingness indicator). In a first set of simulation studies, we used a non-homogeneous Poisson process with rate denoted by $$\lambda(t\mid A(t), M(t), P(t), K; \gamma^{\dagger})= 0.25 ( t+0.05)\text{exp}\left\{\gamma_8 A(t) + \gamma_9 M(t) + \gamma_{10} K_1 + \gamma_{11} K_2 + \gamma_{12} K_3 + \gamma_{13} P(t)\right\}.$$ We tested the four following combinations for the parameters $\gamma^{\dagger}$: $ 1) \gamma^{\dagger}=(0, 0, 0, 0, 0, -5)$ (i.e., no bias due to the visit process expected); 2) $\gamma^{\dagger}=(0.5, 0.3, -0.5, -2, 0, -3)$; 3) $\gamma^{\dagger}=(0.5, -0.5, -0.2, -1, 1, -3)$; and 4) $\gamma^{\dagger}=(-1, -0.8, 0.1, 0.3, -1, -3)$. In a second set of simulation studies, 
we used Bernoulli random variables with probabilities denoted by $\lambda(t\mid A(t), M(t), P(t), K; \gamma)$\\$=\text{expit}\left\{\gamma_1 + \gamma_2 A(t) + \gamma_3 M(t) + \gamma_4 K_1 + \gamma_5 K_2 + \gamma_6 K_3 + \gamma_7 P(t)\right\}$ to simulate the indicators. We tested the four following combinations for the parameters $\gamma$: $ 1)       
\gamma=(0.4, 0, 0, 0, 0, 0, -5)$ (i.e., no bias due to the visit process expected); 2) $\gamma=(0.4, 1, -1, -0.5, -2, 0, -3)$;\\ 3) $\gamma=(0.4, 0.5, -0.5, -0.2, -1, 1, -3)$; and 4) $\gamma=(0.4, -0.5, 0.8, 0.1, 0.3, -1, -3)$. In both sets of simulations, the observation indicators could be obtained by simulating Bernoulli random variables with probability proportional to the Poisson rate or the Bernoulli probability. When, at a given time point, the simulated observation indicator was simulated as 0, the corresponding outcome value was set to missing.

The ordinary least squares estimator, the inverse probability of treatment-weighted estimators and the FIPTM estimators were obtained by fitting standard linear regressions (e.g., with the function \texttt{lm} in R) using the weight statement whenever applicable for the inverse probability of treatment or the inverse intensity of visit weights. The conditional outcome mean models were obtained by fitting two different linear models, one conditioning on the confounders and the treatment (corresponding to $\mu_a\{K, A=a; \hat{\alpha}_K \}$ in the main manuscript) and one conditional on the visit predictors and the treatment (corresponding to $\mu_a \{V, A=a; \hat{\alpha}_V \} $). 

To assess the performance of the AAIIW estimator when the model $\mu_a \{K, A=a;  \alpha_K \}$ is correctly specified, we had to know the true model $\mu_a \{K, A=a; \alpha_{K0} \}$. Based on the outcome generating mechanism, that true model is 
\begin{align*}
\E [Y(t)\mid A(t), K ] &= \E[ \kappa + 1\hspace{0.15cm} A(t) + 0.4\hspace{0.15cm} K_1 + 0.05 \hspace{0.15cm} K_2 - 0.6 \hspace{0.15cm} K_3   \\ 
&\hspace{1cm} +3 \left\{ M(t) - E[M(t)\mid A(t), K_1, K_2, K_3] \right\}  + 0.3\hspace{0.15cm} P(t)  + \epsilon(t) \mid A(t), K] \\
& = \kappa + A(t)  + 0.4  K_1+ 0.05 K_2 - 0.6  K_3  + 0.3 P(t).
\end{align*} 
The variable $P(t)$ is not mandatory to adjust for in our setting, since it should not affect the contrast in the outcome mean across treated ($A(t)=1$) and untreated patients ($A(t)=0$) (unless there was an interaction term between $P(t)$ and the treatment in the outcome generating mechanism). The correctly specified model $\mu_a \{V, A=a; \alpha_V\}$ merely contains all the predictors that were used in generating the outcome above (i.e., the treatment, the re-centered mediator, the pure predictor, and all the confounders).

\newpage

\noindent \textbf{Supplementary Material D} \vspace{0.2cm}\\

\noindent \textbf{Additional results of the simulation studies: sample of size 250 with the use of a nonhomogeneous Poisson rate, or the use of a Bernoulli probability (sample sizes 250 and 1000) to simulate the observation indicators and Monte Carlo empirical bias and mean square error (MSE) in the simulation studies}

\setcounter{figure}{0}
\renewcommand{\figurename}{Suppl. Fig.}
\setcounter{table}{0}
\renewcommand{\tablename}{Suppl. Table}

 \begin{figure}[H]
\begin{center}
\includegraphics[width= 0.85\textwidth]{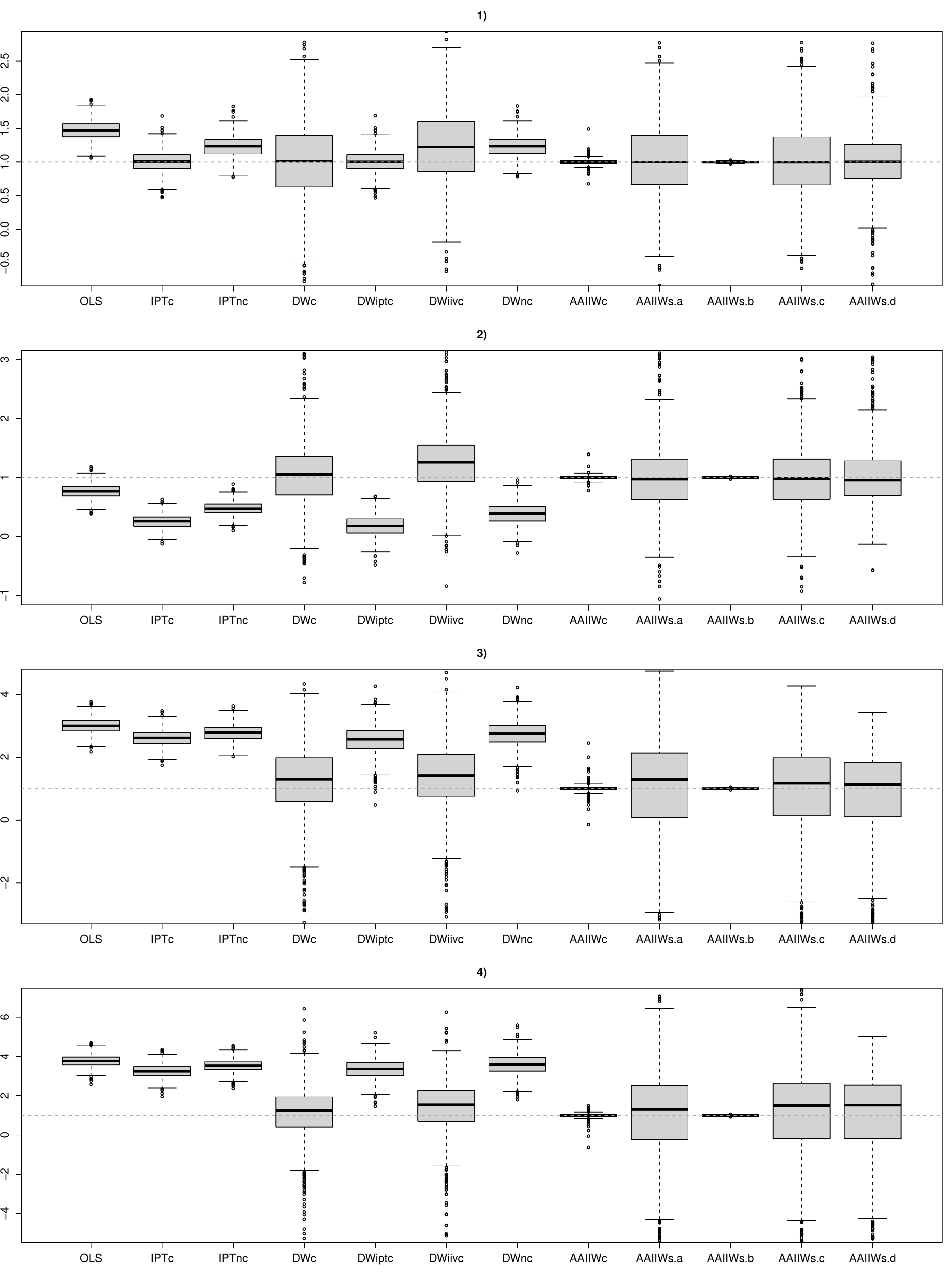}   
 \caption{Results of the simulation studies with a sample size of 250 using a nonhomogeneous Poisson rate to simulate the observation indicators and the Andersen and Gill model with Breslow estimator to estimate the IIV weights. Each boxplot represents the distribution of 1000 estimates for the corresponding estimator. The dashed line represents the gold standard, i.e., the true value for the marginal effect of exposure that equals to 1. Different strengths of the visit process on covariates are represented with scenarios 1) $\gamma=(0, 0, 0, 0, 0, -5)$ (i.e., no bias due to the visit process expected); 2) $\gamma=(0.5, 0.3, -0.5, -2, 0, -3)$; 3) $\gamma=(0.5, -0.5, -0.2, -1, 1, -3)$; and 4) $\gamma=(-1, -0.8, 0.1, 0.3, -1, -3)$.}\label{fig7}
\end{center}
\end{figure}

\begin{figure}[H]
\begin{center}
\includegraphics[width= 0.9\textwidth]{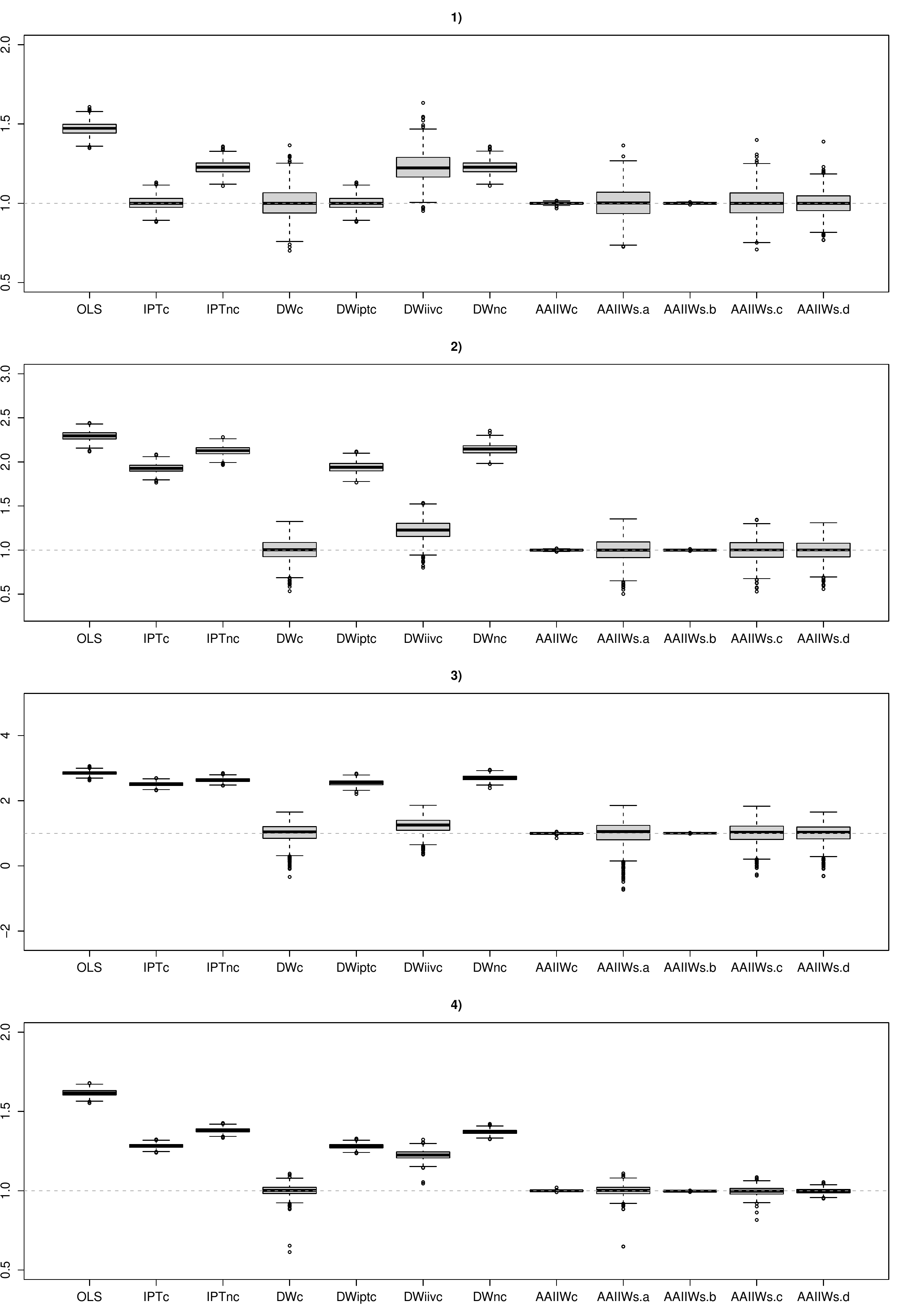}   
 \caption{Results of the simulation studies using Binomial probabilities with a sample size of 1000. Each boxplot represents the distribution of 1000 estimates for the corresponding estimator. The dashed line represents the gold standard, i.e., the true value for the marginal effect of exposure that equals to 1. Different strengths of the visit process on covariates are represented with scenarios 1) $\gamma=(0.4, 0, 0, 0, 0, 0, -5)$ (i.e., no bias due to the visit process expected); 2) $\gamma=(0.4, 1, -1, -0.5, -2, 0, -3)$; 3) $\gamma=(0.4, 0.5, -0.5, -0.2, -1, 1, -3)$; and 4) $\gamma=(0.4, -0.5, 0.8, 0.1, 0.3, -1, -3)$.}\label{fig8}
\end{center}
\end{figure}

\begin{figure}[H]
\begin{center}
\includegraphics[width= 0.9\textwidth]{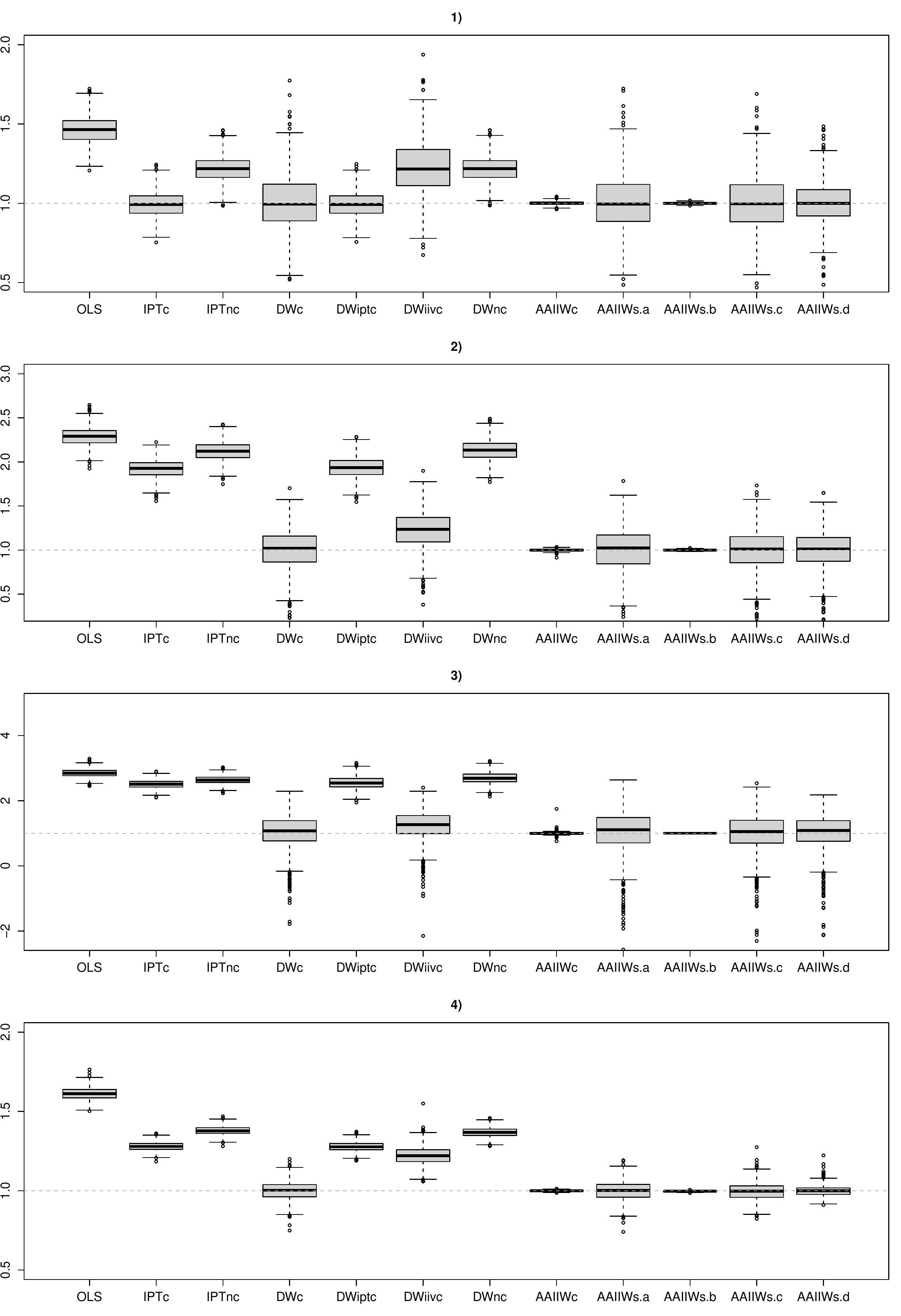}   
 \caption{Results of the simulation studies using Binomial probabilities with a sample size of 250. Each boxplot represents the distribution of 1000 estimates for the corresponding estimator. The dashed line represents the gold standard, i.e., the true value for the marginal effect of exposure that equals to 1. Different strengths of the visit process on covariates are represented with scenarios 1) $\gamma=(0.4, 0, 0, 0, 0, 0, -5)$ (i.e., no bias due to the visit process expected); 2) $\gamma=(0.4, 1, -1, -0.5, -2, 0, -3)$; 3) $\gamma=(0.4, 0.5, -0.5, -0.2, -1, 1, -3)$; and 4) $\gamma=(0.4, -0.5, 0.8, 0.1, 0.3, -1, -3)$.}\label{fig9}
\end{center}
\end{figure}

 \begin{table}[H]
\caption{Simulations results, empirical bias and mean squared error (MSE) }

 \begin{center}
 
\begin{tabular}{|l|c|c|c| c|c|c|  c| c|c|c|} 
\hline  
            & Average $N(\tau)$&  Gamma set&      \multicolumn{4}{|c|}{Poisson rate} &\multicolumn{4}{|c|}{Bernoulli probability} \tabularnewline 
         Estimator  &  $(A_i(t)=0, 1)$  & $\gamma$    & \multicolumn{2}{|c|}{$n=1000$}   &  \multicolumn{2}{|c|}{$n=250$}  & \multicolumn{2}{|c|}{$n=1000$}   &  \multicolumn{2}{|c|}{$n=250$}  
        \tabularnewline  
        &  Poisson $\dagger$ & Poisson $\dagger$ & Bias& MSE  & Bias& MSE & Bias& MSE  & Bias& MSE\tabularnewline 
        &  or Bernoulli $\ddagger$ & or Bernoulli $\ddagger$  & & && & & & & \tabularnewline 
        \hline

       OLS & (12, 12)$^\dagger$ & 1$^\dagger$ & 0.47 & 0.22 & 0.47& 0.24 & 0.47& 0.22 & 0.46& 0.22 \tabularnewline
       IPTc & (18, 18)$^\ddagger$&1$^\ddagger$ & $<$0.01 & 0.01 &0.01& 0.03& $<$0.01& $<$0.01 & 0.01& 0.01\tabularnewline
       IPTnc & & & 0.23 & 0.06& 0.23& 0.08& 0.23 & 0.05 & 0.22 & 0.05\tabularnewline
       DWc & & & 0.01& 0.13& 0.01& 0.44 & $<$0.01 & 0.01 & $<$0.01 & 0.03\tabularnewline
       DWiptc & & &  $<$0.01& 0.01 & 0.01& 0.03& $<$0.01 &$<$0.01 & 0.01& 0.01\tabularnewline
       DWiivc & & & 0.23 & 0.18 &0.22& 0.47& 0.23 & 0.06 & 0.22& 0.08\tabularnewline
       DWnc & & & 0.23 & 0.06&0.23& 0.08 &  0.23& 0.05& 0.22 & 0.05\tabularnewline
       AAIIWc & & & $<$0.01 & $<$0.01 &$<$0.01& $<$0.01& $<$0.01& $<$0.01& $<$ 0.01 &$<$ 0.01 \tabularnewline
       AAIIWs.a. & & & 0.01 & 0.15 & $<$0.01& 0.73&   $<$0.01 & 0.01  & $<$0.01 & 0.03\tabularnewline
       AAIIWs.b. & & & $<$0.01 &  $<$0.01 & $<$0.01&$<$0.01& $<$0.01 &$<$0.01 & $<$0.01&$<$0.01\tabularnewline
       AAIIWs.c. & & &$<$0.01 & 0.14&$<$0.01& 0.84 &$<$0.01 & 0.01 & $<$0.01& 0.03\tabularnewline
       AAIIWs.d. & & & $<$0.01& 0.10 &$<$0.01& 0.46&  $<$0.01& 0.01 & $<$0.01& 0.02\tabularnewline \hline
        OLS & (22, 17)$^\dagger$ & 2$^\dagger$ & 0.22 & 0.05 & 0.23 & 0.07& 1.30 & 1.68 & 1.29 &1.67\tabularnewline
       IPTc & (1, 7)$^\ddagger$ &  2$^\ddagger$ & 0.75& 0.56 &0.75& 0.57& 0.93 & 0.86 &0.92 &0.87\tabularnewline
       IPTnc & & & 0.52 & 0.27 &0.52& 0.29&1.13 & 1.27 &1.12 &1.27\tabularnewline
       DWc & & & 0.05 & 0.08 &0.04& 0.29&$<$0.01  & 0.02 & $<$0.01 & 0.06\tabularnewline
       DWiptc & & &  0.82& 0.68 &0.82& 0.71& 0.94 & 0.89 & 0.93 &0.88\tabularnewline
       DWiivc & & & 0.27 & 0.15 &0.26& 0.37& 0.22 & 0.06 & 0.22 & 0.10\tabularnewline
       DWnc & & & 0.61& 0.38& 0.61& 0.41&1.14  & 1.31&1.13 &1.30 \tabularnewline
       AAIIWc & & & $<$0.01 & $<$0.01&$<$0.01 &$<$0.01& $<$0.01&$<$0.01 & $<$0.01& $<$0.01\tabularnewline
       AAIIWs.a. & & & 0.01 & 0.10&0.02& 0.49&$<$0.01 & 0.02 & $<$0.01& 0.07 \tabularnewline
       AAIIWs.b. & & & $<$0.01 & $<$0.01& $<$0.01&$<$0.01&$<$0.01 &$<$0.01 & $<$0.01&$<$0.01 \tabularnewline
       AAIIWs.c. & & & 0.03 & 0.10 &0.04& 0.51& $<$0.01& 0.02 & $<$0.01& 0.06\tabularnewline
       AAIIWs.d. & & & 0.05 & 0.08 &0.05& 0.42&$<$0.01 & 0.01 & $<$0.01& 0.05\tabularnewline \hline
        OLS & (3, 8)$^\dagger$ & 3$^\dagger$ &  2.02 & 4.09 &2.00& 4.08& 1.85 & 3.43 & 1.85 & 3.43\tabularnewline
       IPTc & (6, 14)$^\ddagger$& 3$^\ddagger$ & 1.62 & 2.63&1.61&2.65 & 1.51 & 2.27 & 1.51&2.28\tabularnewline
       IPTnc & & &  1.79& 3.22 &1.78&3.23& 1.64 & 2.69 & 1.64&2.70\tabularnewline
       DWc & & & 0.01 & 0.67 &0.18& 1.63&   0.01 & 0.09 & 0.04& 0.29\tabularnewline
       DWiptc & & & 1.56 & 2.59 &1.55& 2.62& 1.55& 2.41 & 1.55& 2.44\tabularnewline
       DWiivc & & & 0.21 & 0.57 &0.34&1.55& 0.23 & 0.12 & 0.24& 0.28\tabularnewline
       DWnc & & & 1.77 & 3.17 &1.75& 3.24& 1.70 & 2.91 & 1.70& 2.92\tabularnewline
       AAIIWc & & & $<$0.01 &  $<$0.01&$<$0.01&0.01& $<$0.01 &$<$0.01 & $<$0.01&$<$0.01 \tabularnewline
       AAIIWs.a. & & & 0.38 & 2.68 &0.31& 11.13& $<$0.01 & 0.14 & $<$0.01& 0.91\tabularnewline
       AAIIWs.b. & & &  $<$0.01  &  $<$0.01 & $<$0.01& $<$0.01& $<$0.01 & $<$0.01& $<$0.01&$<$0.01\tabularnewline
       AAIIWs.c. & & & 0.34 & 1.83 &0.28& 9.62& $<$0.01 & 0.10 & 0.01& 0.42\tabularnewline
       AAIIWs.d. & & & 0.45 & 1.80 &0.43& 8.67& $<$0.01 & 0.09 & 0.01& 0.37\tabularnewline \hline
        OLS & (2, 5)$^\dagger$ & 4$^\dagger$ & 2.78 & 7.77&2.78& 7.80 & 0.62 & 0.38 & 0.61 & 0.38\tabularnewline
       IPTc & (81, 58)$^\ddagger$&4$^\ddagger$ & 2.26 & 5.12 &2.25& 5.19& 0.28 & 0.08 & 0.28 & 0.08\tabularnewline
       IPTnc & & & 2.54 & 6.45 &2.53&6.48& 0.38 & 0.15 & 0.38 & 0.14\tabularnewline
       DWc & & & 0.01 & 0.84&0.10&1.93 & $<$0.01 & $<$0.01& $<$0.01& $<$0.01\tabularnewline
       DWiptc & & & 2.35 & 5.57 &2.35&5.79& 0.28& 0.08 & 0.28& 0.08\tabularnewline
       DWiivc & & & 0.25 & 0.91 &0.37&2.08&0.23 & 0.05 & 0.22 & 0.05\tabularnewline
       DWnc & & &  2.59 & 6.78 &2.59&6.98& 0.37 & 0.14 & 0.37 & 0.14\tabularnewline
       AAIIWc & & & $<$0.01 & $<$0.01 &$<$0.01&0.01& $<$0.01 &$<$0.01 & $<$0.01&$<$0.01\tabularnewline
       AAIIWs.a. & & & 0.34 & 4.77 &0.21&13.54& $<$0.01&$<$0.01 &$<$0.01 &$<$0.01\tabularnewline
       AAIIWs.c. & & & 0.38 & 6.53 &0.25&14.89& $<$0.01& $<$0.01& $<$0.01&$<$0.01 \tabularnewline
       AAIIWs.d. & & & 0.64 & 6.33 &0.41&12.58& $<$0.01&$<$0.01 & $<$0.01&$<$0.01\tabularnewline
   \hline
   \end{tabular} 
   
  \scriptsize{$\dagger$. 1: $\gamma$=(0, 0, 0, 0, 0, -5); 2: $\gamma$=(0.5, 0.3, -0.5, -2, 0, -3); 3: $\gamma$=(0.5, -0.5, -0.2, -1, 1, -3); 4: $\gamma$=(-1, -0.8, 0.1, 0.3, -1, -3). \\
  $\ddagger$. 1: $\gamma$=(0.4, 0, 0, 0, 0, 0, -5); 2: $\gamma$=(0.4, 1, -1, -0.5, -2, 0, -3); 3: $\gamma$=(0.4, 0.5, -0.5, -0.2, -1, 1, -3); 4: $\gamma$=(0.4, -0.5, 0.8, 0.1, 0.3, -1, -3).}
\end{center}

\end{table}

\newpage

\noindent \textbf{Supplementary Material E} \vspace{0.2cm}\\
\noindent \textbf{Tables of characteristics in the \textit{Add Health} study  stratified by weighting strategy}

\begin{table}[H]
\caption{Longitudinal characteristics stratified by adolescents receiving counselling therapy or not in the first dataset imputed with multiple imputations by chained equations, \textit{Add Health} study, United States, 1996-2008}

\centering 

\resizebox{\textwidth}{!}{%
\begin{tabular}{|l|c|c|c|c|  } 
\hline 
            & \multicolumn{2}{|c| }{Before inverse probability } &  \multicolumn{2}{|c| }{After inverse probability}  \tabularnewline
            & \multicolumn{2}{|c| }{of treatment-weighting} & \multicolumn{2}{|c| }{of treatment-weighting} \tabularnewline
            \hline
Variable, N (\%) & Counselling & No counselling & Counselling & No counselling \tabularnewline
\hline
\hline
Age, mean (SD) & 20.3 (5.8) & 20.8 (5.7) &20.9 (6.0) & 20.8 (5.7) \tabularnewline
\hline
Female sex& 1579 (60.8)& 11,848 (50.6)  & 13,528 (51.8) & 13,427 (51.6) \tabularnewline
\hline
Weight, mean (SD)& 156.2 (42.6) & 160.7 (44.4) & 160.6 (44.4)& 160.3 (44.3) \tabularnewline
\hline
Socioeconomic status, mean (SD)&6.5 (2.2) & 6.4 (2.2) & 6.4 (2.2)& 6.5 (2.2) \tabularnewline
\hline
Smoking & 1090 (42.0)& 6476 (27.7) & 7451 (28.6) & 7563 (29.1) \tabularnewline
\hline
Depressive mood, mean (SD)$^\dagger$&1.9 (0.9) & 1.4 (0.7)   & 1.8 (0.9)& 1.4 (0.7)\tabularnewline
\hline\end{tabular} }\label{tab11}

\small{$\dagger$. This is not considered as a confounder but as a mediator in our analyses. Therefore, depressive mood was not included in the inverse probability of treatment weights.
Acronym: SD, standard deviation.}

\end{table}

\begin{table}[H]
\caption{Longitudinal characteristics stratified by the alcohol consumption being observed or not in the first dataset imputed with multiple imputations by chained equations and after inducing outcome missingness, \textit{Add Health} study, United States, 1996-2008}

\centering 
\resizebox{\textwidth}{!}{%
\begin{tabular}{|l|c|c|c|c|  } 
\hline 
            & \multicolumn{2}{|c| }{Before inverse intensity} &  \multicolumn{2}{|c| }{After inverse intensity}  \tabularnewline
            & \multicolumn{2}{|c| }{of visit-weighting} &  \multicolumn{2}{|c| }{of visit-weighting}  \tabularnewline
            \hline
Variable, N (\%) & Observed & Not observed & Observed & Not observed \tabularnewline
\hline
\hline
Age, mean (SD) &  20.9 (5.7) & 20.7 (5.7) & 20.9 (5.7) & 21.0 (5.7)  \tabularnewline
\hline
Female sex&   5283 (50.0) & 8144 (52.7)& 2777 (50.2) & 2496 (50.0)\tabularnewline
\hline
Weight, mean (SD)&160.3 (44.3) & 160.2 (45.0) & 159.9 (44.1) & 161.7 (45.0)  \tabularnewline
\hline
Socioeconomic status, mean (SD)&  6.6 (2.1) & 6.5 (2.1) & 6.6 (2.1) & 6.6 (2.1) \tabularnewline
\hline
Smoking & 3159 (29.9) & 4912 (31.8) & 1686 (30.5) & 1412 (28.0)  \tabularnewline
\hline
Depressive mood, mean (SD)& 1.1 (0.4) & 1.6 (0.8) & 1.1 (0.3) & 1.2 (0.4) \tabularnewline
\hline
Counselling &  0.2 (0.4) & 0.1 (0.2) & 0.3 (0.4) & 0.1 (0.3) \tabularnewline
\hline\end{tabular}}\label{tab12}

\small{Acronym: SD, standard deviation.}  
\end{table}

 \end{document}